\documentclass[aps,twocolumn,prl,superscriptaddress,floats,nobibnotes,notitlepage,citeautoscript]{revtex4-1}

\usepackage{graphicx}
\usepackage[utf8]{inputenc}
\usepackage{amsmath,amssymb}
\usepackage{newtxtext}
\usepackage[varvw]{newtxmath}
\usepackage{multirow}
\usepackage{mathtools}
\usepackage[pdftex]{hyperref}

\hypersetup{
    colorlinks,
    linkcolor={blue},
    citecolor={blue},
    urlcolor={blue}
}

\def\<{\langle}
\def\>{\rangle}
\DeclareMathOperator{\Tr}{Tr}

\newcommand{\ve}[1]{\boldsymbol{#1}}

\makeatletter
\def\maketitle{
\@author@finish
\title@column\titleblock@produce
\suppressfloats[t]}

\makeatother

\makeindex

\begin{document}

\title{Metallic and Deconfined Quantum Criticality in Dirac Systems}


\author{Zi Hong Liu}
\affiliation{Institut f\"ur Theoretische Physik und Astrophysik and W\"urzburg-Dresden Cluster of Excellence ct.qmat,\\
Universit\"at W\"urzburg, 97074 W\"urzburg, Germany}
\author{Matthias Vojta}
\affiliation{Institut f\"ur Theoretische Physik and W\"urzburg-Dresden Cluster of Excellence ct.qmat,
Technische Universit\"at Dresden, 01062 Dresden, Germany}
\author{Fakher F. Assaad}
\affiliation{Institut f\"ur Theoretische Physik und Astrophysik and W\"urzburg-Dresden Cluster of Excellence ct.qmat,\\
Universit\"at W\"urzburg, 97074 W\"urzburg, Germany}
\author{Lukas Janssen}
\affiliation{Institut f\"ur Theoretische Physik and W\"urzburg-Dresden Cluster of Excellence ct.qmat,
Technische Universit\"at Dresden, 01062 Dresden, Germany}

\begin{abstract}
Motivated by the physics of spin-orbital liquids, we study a model of interacting Dirac fermions on a bilayer honeycomb lattice at half filling, featuring an explicit global SO(3)$\times$U(1) symmetry.
Using large-scale auxiliary-field quantum Monte Carlo (QMC) simulations, we locate two zero-temperature phase transitions as function of increasing interaction strength.
First, we observe a continuous transition from the weakly-interacting semimetal to a different semimetallic phase in which the SO(3) symmetry is spontaneously broken and where two out of three Dirac cones acquire a mass gap. The associated quantum critical point can be understood in terms of a Gross-Neveu-SO(3) theory.
Second, we subsequently observe a transition towards an insulating phase in which the SO(3) symmetry is restored and the U(1) symmetry is spontaneously broken. 
While strongly 
first order at the mean-field level, the QMC data is consistent with a direct and continuous  transition.
It is thus a candidate for a new type of deconfined quantum critical point that features gapless fermionic degrees of freedom.
\end{abstract}

\maketitle


Metallic quantum criticality corresponds to the spontaneous breaking of a symmetry in a metallic environment triggered by varying a non-thermal control parameter such as doping, magnetic field, or pressure~\cite{PhysRevB.14.1165,PhysRevB.48.7183,moriya2012spin}.
Order-parameter fluctuations often induce non-Fermi-liquid behavior in a temperature-versus-control-parameter window of the phase diagram~\cite{metzner2003soft,PhysRevB.78.035103,2007_HvL_RMP}.
Metallic quantum criticality is pivotal in understanding anomalous transport and strange-metal behavior in strongly correlated materials, such as heavy-fermion compounds~\cite{gegenwart2008quantum} and Cu- and Fe-based high-temperature superconductors~\cite{keimer2015quantum,shibauchi2014quantum}.
In spite of extensive efforts~\cite{PhysRevB.14.1165,PhysRevB.48.7183,moriya2012spin,abanov2003quantum,abanov2004anomalous,metlitski2010quantum1,metlitski2010quantum2,sur2016anisotropic,schlief2017exact,lee2018recent,schlief2018noncommutativity}, a controlled analytical treatment of this problem in the presence of a Fermi surface remains a major challenge.
The main difficulty is to tame the strong quantum fluctuations that arise from the abundance of gapless particle--hole modes near the Fermi surface.
From the numerical point of view \cite{Chubukov2018JC}, the fact that these transitions are characterized by dynamical critical exponents $z>1$ impedes the ability to reach sufficiently low temperatures on large lattices \cite{xu2020identification}.

Dirac systems, in contrast, have emergent Lorentz symmetry: Space and time are interchangeable and $z=1$. Furthermore, instead of an extended Fermi surface, they feature isolated Fermi points.
In the past years, there has been a considerable amount of work investigating dynamical mass generation in these systems. The understanding of such transitions relies on the Gross-Neveu-Yukawa theory, in which an order-parameter field of given symmetry is coupled to a fermion-mass term in the same symmetry sector \cite{zinnjustin1991fourfermion}.
Various instances of these transitions have been studied from the perspective of high-energy~\cite{%
%
gehring15,
%
zerf17,
%
gracey18,
%
iliesiu18,
%
huffman2020fermion}
and solid-state \cite{%
%
herbut06,herbut2009relativistic,janssen14,
%
Assaad13,toldin2015fermionic,otsuka2018quantum,lang2019quantum,liu2019superconductivity,liu2020designer,xu2021competing,liu2021grossneveu} 
physics.
In all of the above examples, the quantum critical points separate Dirac semimetallic states from insulating states with a full gap in the fermionic spectrum. Dirac systems can, however, in principle also support relativistic quantum critical points between two distinct semimetallic phases. This possibility was recently scrutinized in the context of a frustrated spin-orbital model, in which case the fermion degrees of freedom arise from a spin fractionalization mechanism~\cite{seifert2020fractionalized}. If such a transition is realizable, it would represent a Dirac \emph{avatar} of metallic quantum criticality that may be more easily accessible to both numerical and field-theoretical analyses.

In this Letter, we investigate a two-dimensional lattice model of interacting fermions designed to feature such a semimetal-to-semimetal quantum critical point. Inspired by Ref.~\onlinecite{seifert2020fractionalized}, we study the Hamiltonian
\begin{equation}
H=-t\sum_{\left\langle \ve{i},\ve{j} \right\rangle} c_{\ve{i} \sigma \lambda}^{\dagger}c^{\phantom\dagger}_{\ve{j}\sigma \lambda}-J \sum_{\ve{i} \alpha}\left(c_{\ve{i}  \sigma \lambda }^{\dagger}K^{\alpha}_{\sigma \sigma'}\tau^{z}_{\lambda \lambda'}c^{}_{\ve{i} \sigma' \lambda' } \right)^{2},
\label{eq:model}
\end{equation}
where $\langle\ve{i},\ve{j}\rangle$ denote pairs of nearest-neighbor sites of a honeycomb lattice, $\lambda=1,2$ is an additional layer index, and summation over repeated indices is implied. Further, $\sigma=1,2,3$ is an SO(3) flavor index, $(K^{\alpha})_{\sigma \sigma'} = -i \epsilon_{\alpha \sigma \sigma'}$ are the generators of SO(3), and $\tau^{x,y,z}$ are Pauli matrices. The model is particle--hole symmetric such that zero chemical potential corresponds to half filling.

The interaction term in Eq.~\eqref{eq:model} is chosen such that the SO(3) symmetry may be spontaneously broken, resulting in a three-component order parameter $\ve{m} = (m_1,m_2,m_3)^\top$. For a single layer, this leads to the low-energy effective Hamiltonian
\begin{align}
	\label{eq:mass}
	\mathcal H = \Psi^\dagger_\sigma \left[i \gamma_0 \gamma_j p_j \delta_{\sigma\sigma'} + m_\alpha (K^\alpha)_{\sigma\sigma'} \gamma_0\right]\Psi_{\sigma'},
\end{align}
where $i \gamma_0 \gamma_j p_j$ is the usual Dirac Hamiltonian in two spatial dimensions, $j=1,2$. The operator $\ve{K}_{\sigma\sigma'}\gamma_0$ anticommutes with the Dirac Hamiltonian, and as such may be thought of as a mass term, but since $(\ve{m} \cdot \ve{K}) \ve{v} = i \ve{m} \times \ve{v}$ for an arbitrary three-component vector~$\ve{v}$, only two out of three Dirac cones will acquire a gap.
In particular, the wavefunction $\Psi_{\sigma}(\tau,\ve{x}) \equiv m_{\sigma} \psi(\tau,\ve{x})$ will correspond to the massless Dirac fermions. The dynamical generation of such term hence corresponds to metallic quantum criticality in a Dirac system, in contrast to the situation in related models that feature metal-insulator transitions~\cite{liu2019superconductivity,liu21}.
Equation~\eqref{eq:mass} encodes a new class of Gross-Neveu (GN) transitions dubbed GN-SO(3) that have recently been studied using approximate analytical techniques \cite{ray2021fractionalized}.

\begin{figure}[tb]
\centering
\includegraphics[width=\columnwidth]{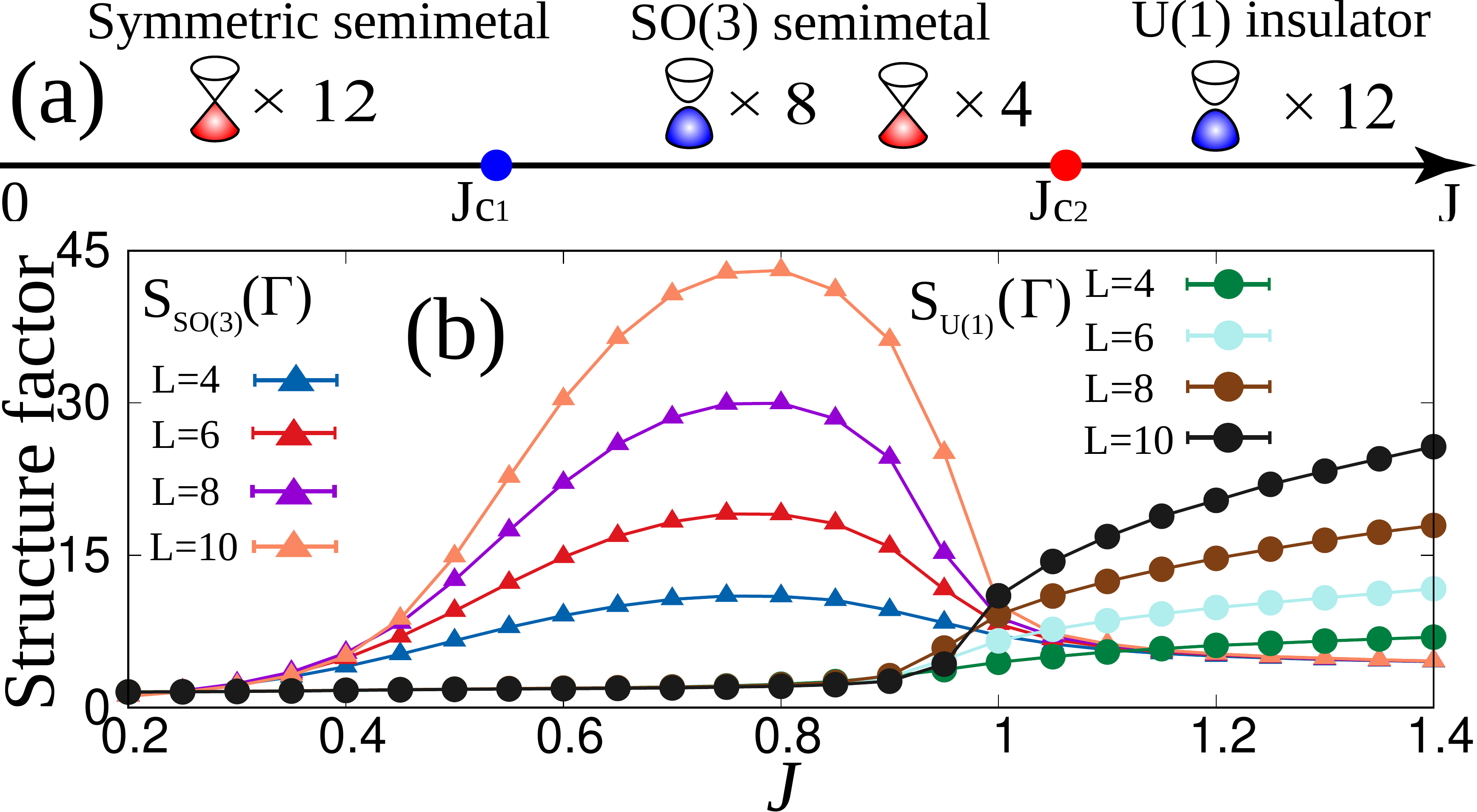}
\caption{(a) Ground-state phase diagram of the model~\eqref{eq:model} as function of interaction strength $J$, obtained from AFQMC simulations.
The semimetal-to-semimetal transition at $J_\mathrm{c1}$ is continuous and can be understood in terms of a GN-SO(3) field theory.
The semimetal-to-insulator transition at $J_\mathrm{c2}$, while strongly first order at the mean-field level, appears continuous. 
(b) Variation of SO(3) and U(1) structure factors $S_\text{SO(3)}(k=\Gamma)$ and $S_\text{U(1)}(k=\Gamma)$ as function of $J$ for different lattice sizes. Extensive structure factors  reflect  spontaneous symmetry  breaking. }
\label{fig:sketch}
\end{figure}

The microscopic model \eqref{eq:model} is amenable to large-scale negative-sign-free auxiliary-field quantum Monte Carlo (AFQMC) simulations~\cite{PhysRevD.24.2278,PhysRevB.31.4403,PhysRevB.40.506}, and Fig.~\ref{fig:sketch} summarizes our key results at zero temperature and half band filling.
In the weakly-interacting limit, the model features a stable semimetallic phase, characterized by $N=12$ irreducible Dirac cones located at the corners K and K$'$ of the first Brillouin zone. At intermediate couplings, a semimetallic SO(3)-symmetry-broken phase with flavor order indeed emerges, in which two thirds of the Dirac cones are gapped out, while one third remains gapless. Remarkably, upon further increasing the interaction strength, we encounter another phase, which now is insulating and displays spontaneously broken U(1) symmetry, corresponding to emergent interlayer coherence.
The first transition is continuous at the mean-field level, and the effects of quantum fluctuations can be understood in terms of the GN-SO(3) field theory studied in~\cite{ray2021fractionalized}.
The second transition is strongly first order at the mean-field level, as is usual for direct transitions between states that break different symmetries. Our numerical results, however, indicate that quantum fluctuations render this order-to-order transition continuous
 (but we cannot exclude it to be weakly first order).  It hence likely represents an example of a new type of deconfined quantum critical point~\cite{senthil04} featuring gapless fermionic degrees of freedom~\cite{zou20}.


\paragraph*{Mean-field analysis.}
Our mean-field approximation relies on the identity
\begin{align}
\label{mean_field.eq}
	-J \sum_{\ve{i} \alpha}\left(c_{\ve{i} }^{\dagger}K^{\alpha}\tau^{z} c_{\ve{i} } \right)^{2} & =
	- J \sum_{\ve{i} \alpha \lambda }  \left( c^{\dagger}_{\ve{i} \lambda} K^{\alpha} c^{\phantom\dagger}_{\ve{i} \lambda}\right)^2    \nonumber \displaybreak[1]\\
	 +2J \sum_{\ve{i}\alpha\sigma \sigma' } & |\epsilon_{\alpha\sigma\sigma'}| \left(\Delta^{\dagger}_{\ve{i}\sigma}\Delta^{\phantom\dagger}_{\ve{i}\sigma'}  +
	n^{\dagger}_{\ve{i}\sigma} n^{\phantom\dagger}_{\ve{i}\sigma'}  \right) 
\end{align}
with $\Delta^{\dagger}_{\ve{i}\sigma} = c^{\dagger}_{\ve{i}\sigma 1} c^{\dagger}_{\ve{i}\sigma 2}  $   and  $n^{\dagger}_{\ve{i}\sigma} = c^{\dagger}_{\ve{i}\sigma 1} c^{\phantom\dagger}_{\ve{i}\sigma 2} $.
The above allows us to define an SO(3) order parameter for staggered flavor order, $m_{\alpha}/2 = (-1)^{\ve{i}} \langle c^{\dagger}_{\ve{i} \lambda} K^{\alpha} c^{\phantom\dagger}_{\ve{i} \lambda} \rangle $, and a U(1) order parameter for interlayer coherence, $ V/2 = (-1)^{\ve{i}} \langle n^{\dagger}_{\ve{i}\sigma}  \rangle$.
In the spirit of the continuum limit of Eq.~\eqref{eq:mass}, the order parameters map onto $\Psi^{\dagger}_{\sigma \lambda}	\gamma_0   K^{\alpha}_{\sigma \sigma'}  \Psi^{\phantom\dagger}_{\sigma' \lambda} $  and $ \Psi^{\dagger}_{\sigma \lambda}	\gamma_0  \tau^{x}_{\lambda\lambda'}  \Psi^{\phantom\dagger}_{\sigma \lambda'} $, respectively, and open partial and full gaps in the fermion spectrum. Since in the QMC calculations we have not observed  superconductivity, we omit the corresponding term in the mean-field approximation.   For details of the calculations, see~\cite{supplemental}. 

Figure~\ref{fig:mftphasediagram}(a) shows the mean-field order parameters as function of $J/t$. The symmetric Dirac  phase has a low-energy density of states $N(\omega)  =  \alpha N | \omega| $ that  changes to  $N(\omega)  =  \alpha \frac{N}{3} | \omega| $ in the SO(3)-broken phase, consistent with two out of three Dirac cones acquiring a mass gap, Figs.~\ref{fig:mftphasediagram}(b,c). At larger values of $J/t$,  we observe a strong first-order transition to a U(1)-broken state whose fermion spectrum is gapped, Fig.~\ref{fig:mftphasediagram}(d).

\begin{figure}[tb]
\centering
\includegraphics[width=\columnwidth]{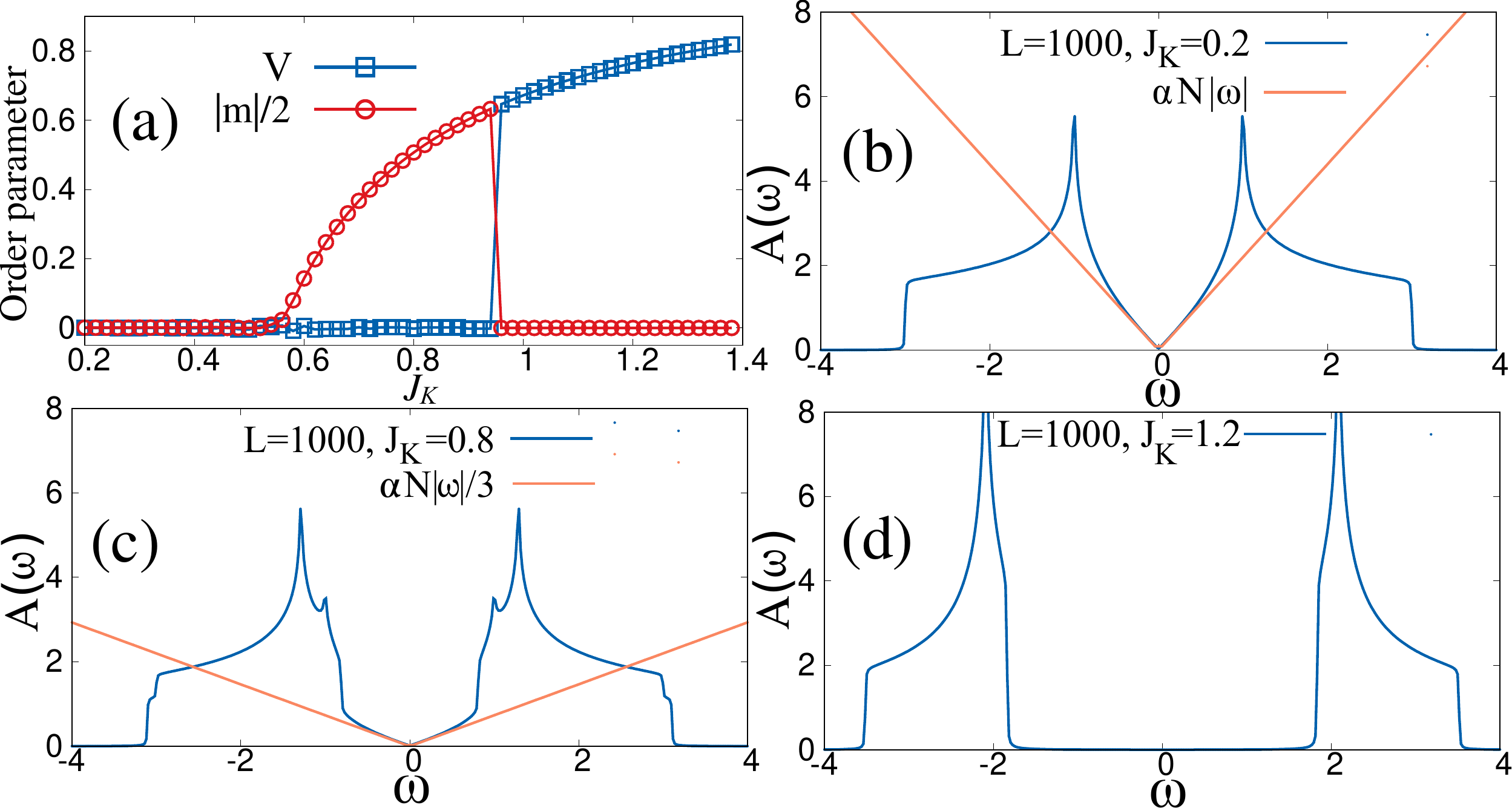}
\caption{
Mean-field results for the model \eqref{eq:model}, obtained on an $L=60$ lattice.
(a) SO(3) and U(1) order parameters $m_{\alpha}$ and $V$ as function of $J$.
(b,c,d) Single-particle density of states for representative fixed values of $J$ in the three phases.
}
\label{fig:mftphasediagram}
\end{figure}


\paragraph*{QMC simulations.}

We have used the ALF  \cite{ALF_v2} implementation of AFQMC, and employed
the finite-temperature grand canonical  and  projective approaches. We used a symmetric Suzuki-Trotter decomposition to control the systematic error in observables~\cite{suzuki1976generalized,trotter1959product} and adopted an imaginary time step $\Delta\tau t = 0.05$ for the finite-temperature algorithm and $\Delta\tau t =  0.1$ for  the projective formulation.

%

The absence of a negative-sign problem for positive values of $J$ stems from a particular time-reversal symmetry that relies on the bilayer structure of the model. After a 
Hubbard-Stratonovich (HS) decomposition  of the perfect-square  interaction term,  the resulting one-body  Hamiltonian,  for a given  space-time  configuration of HS
 fields, has a time-reversal symmetry, defined as $T^{-1}   z  c^{\dagger}_{\ve{i} \sigma \lambda} T^{}  =   \overline{z}  i \tau^{y}_{\lambda \lambda'}c^{\dagger}_{\ve{i} \sigma \lambda'}$. This stems from the fact that the SO(3) generators are purely imaginary. Hence, the eigenvalues of the  fermion matrix  occur in complex conjugate pairs such that positivity of the determinant follows \cite{Wu04}.
To minimize size effects we follow   Ref.~\cite{Assaad01}  and thread the lattice  with a magnetic  flux quantum of opposite sign in  the  two layers \cite{supplemental}.  
It is  interesting to note that introducing a chemical potential will
not  break  this  time-reversal symmetry,  and  simulations at finite  doping  are amenable  to negative-sign-free QMC.


\paragraph*{QMC results.}
We carry out QMC simulations of the model~\eqref{eq:model}  on $L=6,9,12,15,18$  lattices with $6L^2$ orbitals per honeycomb layer, set $t=1$, and scan as function of $J$.   Our results are  summarized  schematically  in Fig.~\ref{fig:sketch}(a).    Each  phase  is   characterized by  spontaneous symmetry breaking  
 and   diverging  structure factor  at   $\ve{Q}=\Gamma$, Fig.~\ref{fig:sketch}(b).  To detect SO(3) symmetry breaking we consider 
 $S_{\text{SO(3)}}(\ve{k},\tau)=\sum_{\ve{r}_{ij},\lambda}e^{-i\ve{k}\cdot\ve{r}_{ij}} \langle c_{\ve{i},\lambda}^{\dagger}(\tau)\ve{K}c_{\ve{i},\lambda}\left(\tau\right)\cdot c_{\ve{j},\lambda}^{\dagger}\left(0\right)\ve{K} c_{\ve{j},\lambda}\left(0\right)\rangle$ 
 and   for the U(1) phase,
 $S_{\text{U(1)}}(\ve{k},\tau)=\frac{1}{2}\sum_{\ve{r}_{ij}\sigma}e^{-i\ve{k}\cdot\ve{r}_{ij}}\langle n_{\ve{i},\sigma}^{\dagger}(\tau)n_{\ve{j},\sigma}^{\phantom{\dagger}}+n_{\ve{j},\sigma}^{\phantom{\dagger}}(\tau)n_{\ve{i},\sigma}^{\dagger}\rangle$.
Here, $\ve{r}_{ij}$ corresponds to the distance between  the unit cells of $\ve{i}$ and $\ve{j}$. 

Each  phase has a distinct signature in the single-particle spectral function $A(\ve{k},\omega)$. We extract this quantity from the ground-state imaginary-time-displaced fermion Green's functions $G(\ve{k},\tau)   =\frac{1}{\pi}\int d \omega e^{-\tau\omega} A(\ve{k},\omega)$ by using the   ALF~\cite{ALF_v2}  implementation of the stochastic analytic continuation method~\cite{beach2004identifying}.  
In  Fig.~\ref{fig:qmcakw}(b), in the symmetric phase, the fermion spectrum reveals semimetallic behavior. In Fig.~\ref{fig:qmcakw}(c), in the SO(3)-breaken phase, part of low-energy spectral weight is removed, but a finite weight at the Dirac point is still apparent. To demonstrate this  explicitly, we  make use of the fact that for a gapless  mode,  the quasiparticle residue reads $Z(\ve{k})=2\Tr G(\ve{k},\beta/2)$~\cite{brunner2000single}.  We use a  $\beta=L$ scaling and extrapolate $Z$ to the thermodynamic limit, see Fig.~\ref{fig:qmcakw}(a). In the semimetallic phase, the quasiparticle residue extrapolates to the free Dirac-metal value $Z(k\!=\!K) = 6$. In the SO(3)-symmetry-breaking phase, using a polynomial fit, we obtain the estimated quasiparticle residue $Z(k\!=\!K)=1.9(1)$. The ratio of the quasiparticle weights is close to three,  as expected from the gapping out of 2/3 of the Dirac cones. Finally in the U(1)-broken phase, the spectrum shows a full gap, Fig.~\ref{fig:qmcakw}(d).

\begin{figure}[tb]
\centering
\includegraphics[width=\columnwidth]{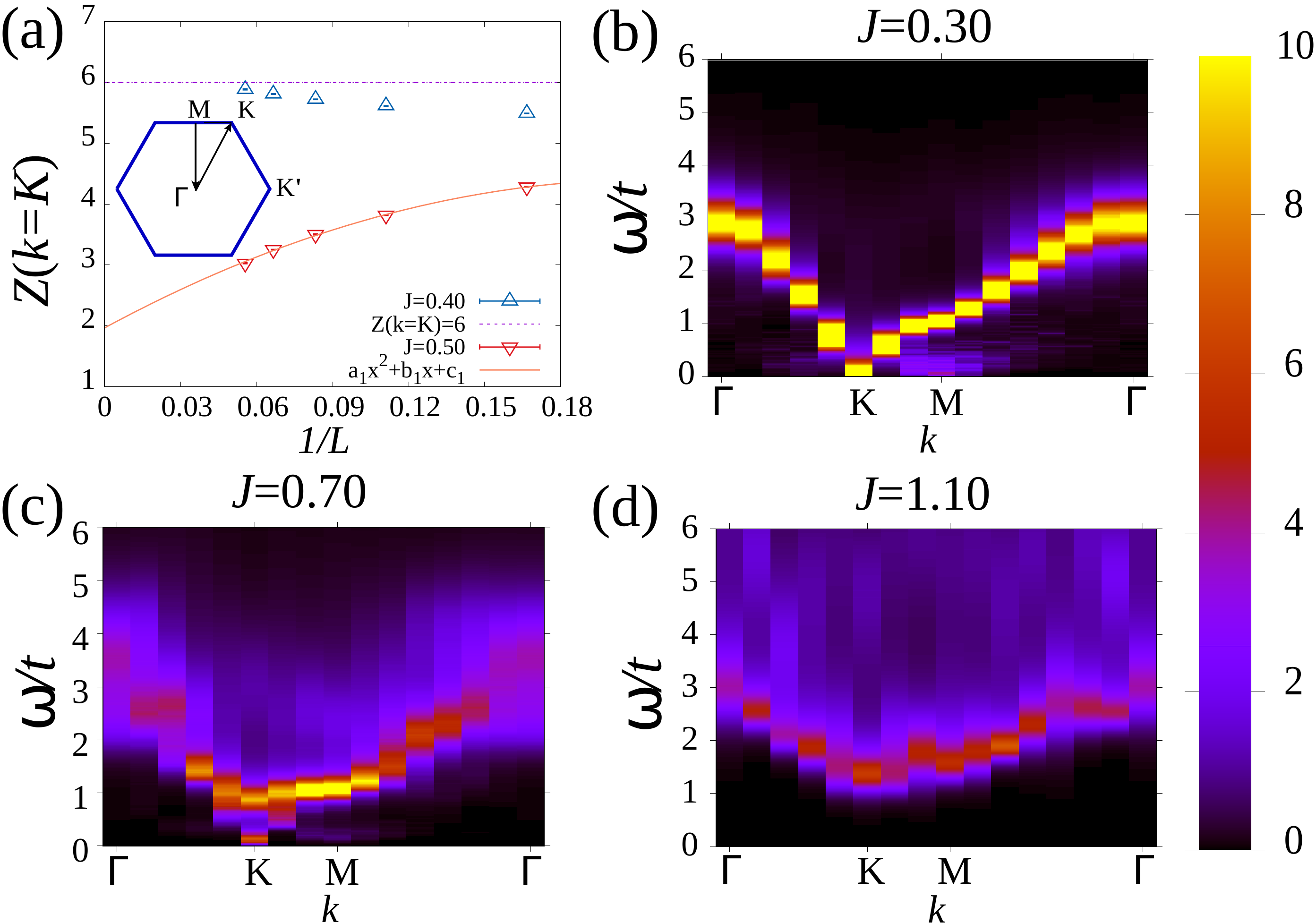}
\caption{
(a) Quasiparticle weight $Z(k\!=\!K)$ as a function of inverse system size, $1/L$, in the symmetric ($J=0.40$) and SO(3)-broken ($J=0.50$) phases, respectively. The polynomial fitting of the second curve yields $Z=1.9(1)$ for $L \to \infty$.
(b-d) Fermion spectral function $A(\ve{k},\omega)$ along the path in momentum space shown in the inset of (a), shown for (b) the symmetric phase at $J=0.30$, (c) the SO(3)-broken phase at $J=0.70$, and (d) the U(1)-broken phase at $J=1.10$. Finite weight at the Dirac point is visible in (b) and (c), corresponding to semimetallic behavior, albeit with a reduced low-energy weight in the case of the SO(3)-broken semimetal (c). The low-energy  weight  at the $M$ point    especially   visible in (b) is an artifact of the magnetic flux and does not survive the thermodynamic limit~\cite{supplemental}.}
\label{fig:qmcakw}
\end{figure}

The two phase transition points are located by monitoring the renormalization-group (RG) invariant correlation ratio \cite{Kaul15}
$R_\mathrm{c}=1-\frac{S\left(\ve{k}=\ve{Q}+d\ve{k},\tau=0\right)}{S\left(\ve{k}=\ve{Q},\tau=0\right)}$, where $S\left(\ve{k},\tau=0\right)$ is the structure factor of either the SO(3) or the U(1) order, $\ve{Q}$ is the ordering  wavevector and $d\ve{k}$ the smallest momentum on the considered lattice.  

To investigate the first phase transition,  we will assume  $z=1$ and adopt a $\beta = L$ scaling  within the finite-temperature AFQMC  algorithm.
As apparent in Fig.~\ref{fig:Rc_collapse}(a), this phase transition involves only SO(3) symmetry  breaking since a clear crossing is observed in $R_\mathrm{c}^\text{SO(3)}$. On the other  hand, $R_\mathrm{c}^\text{U(1)}$   vanishes for increasing system size, thus excluding long-range U(1) order in the  considered parameter range.
\begin{figure}[tb]
\centering
\includegraphics[width=\columnwidth]{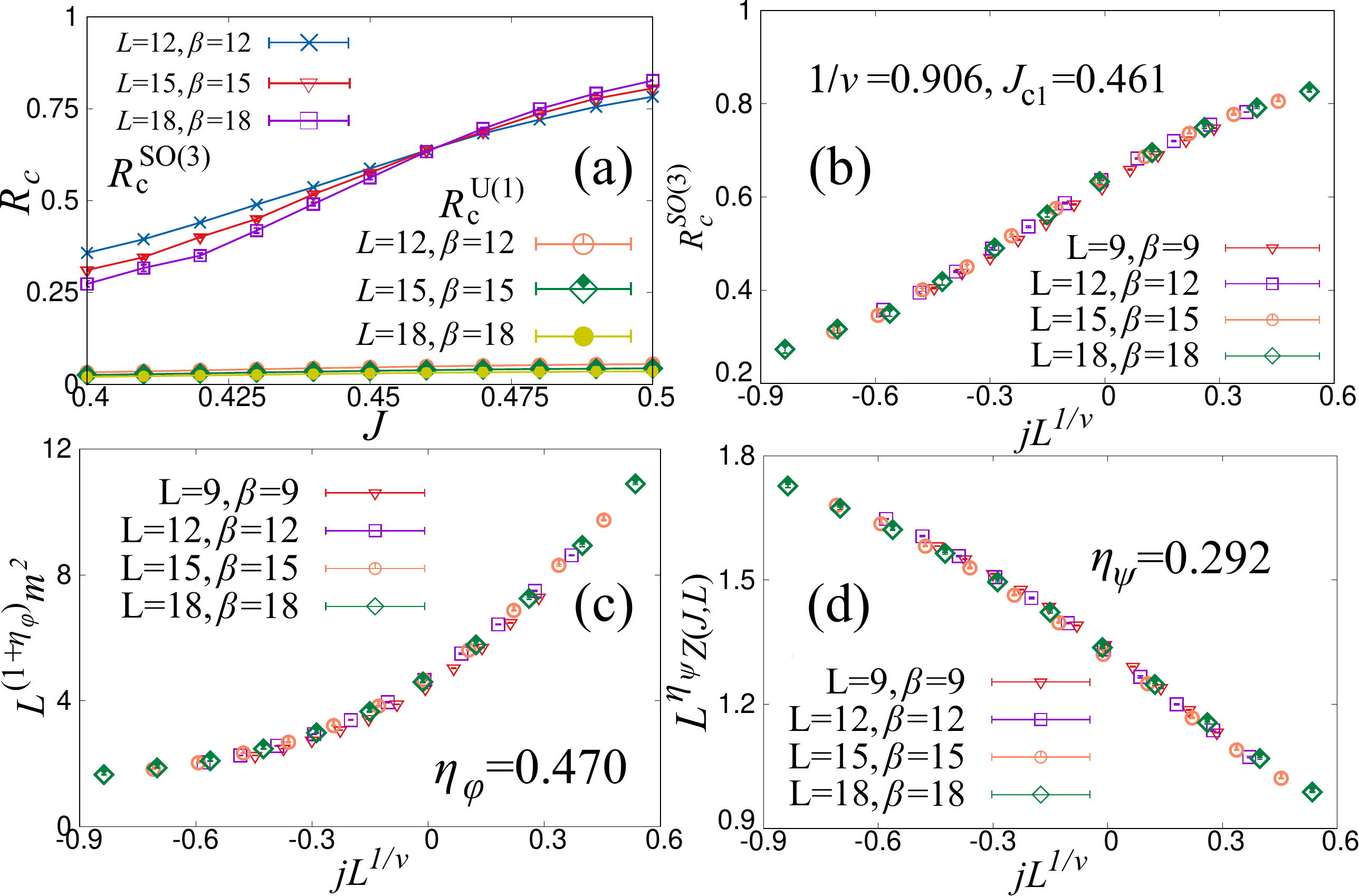}
\caption{
QMC characterization of GN-SO(3) transition at $J_\mathrm{c1}$.
(a)~Correlation ratios of the  U(1) and SO(3) order parameters.
\mbox{(b-d)}:~Scaling collapse near $J_\mathrm{c1}=0.461$ of (b)~correlation ratio $R_\mathrm{c}^\text{SO(3)}$, (c)~order parameter $m^2$ and (d)~fermion quasiparticle weight $Z$, as function of $j L^{1/\nu}$, with $j=J-J_\mathrm{c1}$ and $\nu$ the correlation-length exponent.}
\label{fig:Rc_collapse}
\end{figure}
In the quantum critical region, we expect the correlation ratio $R^\text{SO(3)}_\mathrm{c}$ to obey the finite-size scaling (FSS) ansatz~\cite{campostrini2014finite} $R^\text{SO(3)}_\mathrm{c}(J,L)  =f_{0}^{R}(jL^{1/\nu}) + L^{-\omega}f_1^R(jL^{1/\nu})$, where $j=J-J_\mathrm{c1}$. $f_0^R$ and $f_1^R$ are  scaling functions and $\omega$ is
 the leading-correction-to-scaling exponent. To extract the values of the critical exponent $\nu$ and the critical point $J_\mathrm{c}$, we fit the scaling function $f_0^{R}(jL^{1/\nu})$ to a polynomial~\cite{supplemental}.
In our simulations, the crossing point in  the $R^\text{SO(3)}_\mathrm{c}$ data  becomes size independent  within our  accuracy.  Hence for  $L\geq 12$, corrections to scaling can be omitted.   Our results are consistent with  $1/\nu=0.906(35)$ and $J_\mathrm{c1}=0.461(1)$. The data collapse of $R^\text{SO(3)}_\mathrm{c}(J,L)$  is depicted in  Fig.~\ref{fig:Rc_collapse}(b).
The  bosonic, $\eta_{\phi}$,  and fermionic, $\eta_{\psi}$, anomalous  dimensions are related to the FSS ansatz of the SO(3) order parameter $m^{2}=S_\text{SO(3)}(\mathbf{Q},\tau\!=\!0)/L^{2}$ and  $Z(J,L)=G(J,L)/G(0,L)$, where $G(J,L)=\frac{1}{6}\sum_{\sigma,\lambda}\langle c^{\dagger}_{\ve{0} \sigma \lambda}(\beta/2)c_{\ve{0} \sigma \lambda}(0)\rangle$ at  interaction strength $J$. At the critical point,  and neglecting corrections to scaling,  these two quantities scale as $m_\text{SO(3)}^{2}(j,L)=L^{-(1+\eta_{\phi})}f^{m}(jL^{1/\nu})=L^{-(1+\eta_{\phi})}\tilde{f}^{m}(R_\mathrm{c}^\text{SO(3)}(J,L))$ and $Z(J,L) = L^{-\eta_{\psi}}f^{z}(jL^{1/\nu}) = L^{-\eta_{\psi}}\tilde{f}^{z}(R^\text{SO(3)}_\mathrm{c}(J,L))$ \cite{liu2019superconductivity}. 
 Here we use the correlation ratio $R^\text{SO(3)}_\mathrm{c}(J,L)$ as a dimensionless quantity to replace the variable $jL^{1/\nu}$ so as to reduce the number of fit variables. Following the regression result of the scaling function, we obtain the estimates $\eta_{\phi}=0.470(13)$ and $\eta_{\psi}=0.292(10)$.  Figures~\ref{fig:Rc_collapse}(c,d) show the data collapse  using the estimated exponents.  Consistent  results are obtained   when collapsing the data with respect to $jL^{1/\nu}$ with $\nu$~\cite{supplemental}.

At the mean-field level, the transition between the SO(3)-broken and U(1)-broken states is  strongly first order. However, the QMC results for $R_\mathrm{c}^\text{SO(3)}$ and $R_\mathrm{c}^\text{U(1)}$ do not suggest a strong first-order transition, Figs.~\ref{fig:u1_so3_t0.pdf}(a,b). To assess if there is a coexistence regime, we carry out a crossing-point analysis to  determine the coupling at which the  SO(3) [U(1)] order is suppressed (appears).
We determine  the finite-size critical couplings $J^\text{SO(3)/U(1)}_\mathrm{c2}(L)$ by $R_\mathrm{c}(J_\mathrm{c2}(L),L)=R_\mathrm{c}(J_\mathrm{c2}(L),L+3)$. As the system size $L\rightarrow \infty$, the finite-size critical couplings $J^\text{SO(3)/U(1)}_\mathrm{c2}(L)$ scale as $J_\mathrm{c2}^\text{SO(3)/U(1)} +aL^{-e}$, where $e=1/{\nu}+\omega$ and $a$ is a nonuniversal constant. The results, plotted in Fig.~\ref{fig:u1_so3_t0.pdf}(c), suggests that within our  accuracy, $J_\mathrm{c2}^\text{SO(3)} = J_\mathrm{c2}^\text{U(1)}$.  In Fig.~\ref{fig:u1_so3_t0.pdf}(d), we plot the first derivative of the free energy with respect to $J$. Within our accuracy, we do not observe a  discontinuity expected for a first-order transition.    Consistent estimates of the correlation-length exponent from the U(1) and SO(3)     structure  factors are reported in  \cite{supplemental}.

\begin{figure}[tb]
\centering
\includegraphics[width=\columnwidth]{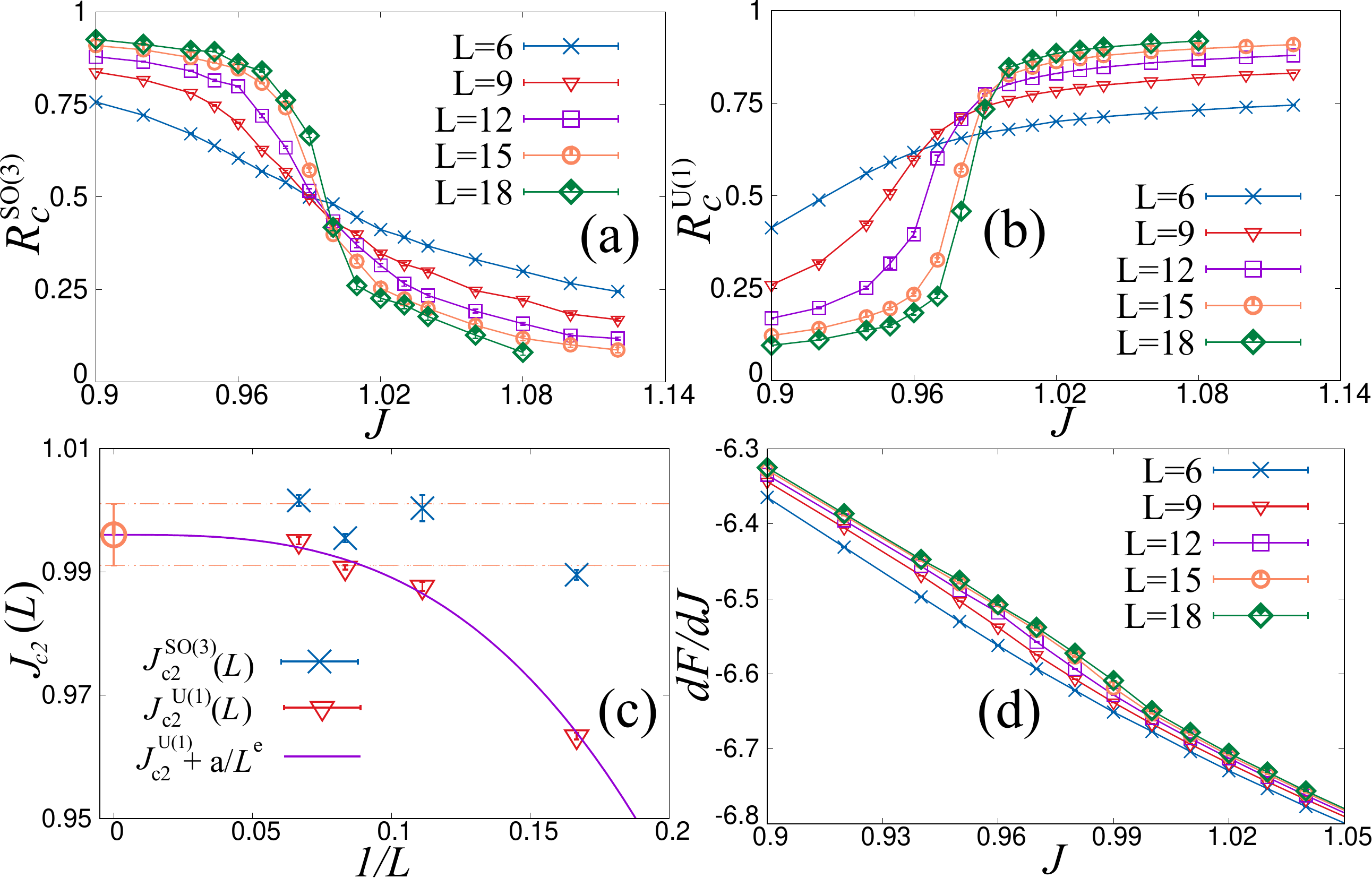}
\caption{
QMC characterization of SO(3)-U(1) transition at $J_\mathrm{c2}$.
(a,b)~Correlation ratios as function of $J$ across transition.
(c)~Finite-size critical couplings $J^\text{SO(3)/U(1)}_\mathrm{c2}(L)$ as function of $1/L$. While $J_\mathrm{c2}^\text{U(1)}(L)$ increases with system size, $J_\mathrm{c2}^\text{SO(3)}(L)$ stabilizes within our accuracy as $L\ge 9$. By extrapolating $J_\mathrm{c2}^\text{U(1)}(L)$ using the power-law ansatz $J_\mathrm{c2}+a/L^e$, we obtain the estimate $J_\mathrm{c2}=1.0013(18)$. 
(d)~First derivative of free energy $dF/dJ$ near $J_\mathrm{c2}$, showing no discontinuity within our accuracy.}
\label{fig:u1_so3_t0.pdf}
\end{figure}


\paragraph*{Discussion and summary.}
We  have introduced a model Hamiltonian, amenable to large-scale negative-sign-free QMC simulations, that supports metallic quantum criticality in Dirac systems. The SO(3) order generates mass in two out of three Dirac cones. Using a FSS analysis, we  estimate the critical exponents of the SO(3)-ordering transition
\begin{align}
1/\nu=0.906(35),\ \eta_{\phi}=0.470(13),\ \eta_{\psi}=0.292(10)
\end{align}
for the correlation-length exponent and the boson and fermion anomalous dimensions. We expect our model to fall into the GN-SO(3) universality class, with $N=12$ two-component Dirac fermions. Calculations in~\cite{ray2021fractionalized}, based on $\epsilon$ expansion, large-$N$, and functional renormalization-group  approaches, yield exponents that differ from our estimates: $1/\nu=0.93(4)$, $\eta_{\phi}=0.83(4)$, $\eta_{\psi}=0.041(12)$. 
 While we  cannot exclude that  the   discrepancy   stems from finite-size effects  in the QMC or  convergence issues in the analytical approaches,  they are large enough to    speculate if topological defects in the field configurations---not included in~\cite{ray2021fractionalized}---play a role, see below.
Note that within the QMC approach one can in principle systematically carry out calculations at $N=12n$ and thereby test the validity of the large-$N$ approach.

At larger couplings, the model   shows an order-to-order  transition between  the  SO(3) semimetal and a U(1) insulator. While at the mean-field level this transition is  strongly  first order,  the  QMC data on lattice sizes with up to $18 \times18$ unit cells can be interpreted in  terms  of a continuous and direct transition, thereby  providing    an instance of a new type of deconfined quantum critical point~\cite{senthil04} featuring gapless fermionic degrees of freedom~\cite{zou20}. Clearly, we cannot  exclude the possibility of a weakly-first-order transition,  in which, as suggested in the realm of deconfined criticality,  the correlation length  grows  beyond  the  accessible system size  due to the proximity of a critical fixed point in the complex-coupling plane \cite{Nahum15,wang17,serna19,Nahum19,WangC19,gorbenko18a,gorbenko18b}. 

The  discrepancy  between the QMC  critical exponents  and  those of  Ref.~\onlinecite{ray2021fractionalized},  as  well as  the  seemingly-continuous transition between the  SO(3)-broken semimetal and  U(1)-broken insulator  leads us to ask the  question if the topology  of the SO(3) order  parameter can play a role. For a  given SO(3) mass  term,   $[\ve{m}(\ve{x}) \cdot \ve{K}]  \gamma_0$,  the  wavefunction of the gapless  Dirac cone    reads  $\Psi_{\sigma \lambda}(\ve{x})   =  m_{\sigma}(\ve{x}) \psi_{\lambda}(\ve{x})$.  Consider an  interface where  on both sides  the  vectors $\ve{m}$   are  orthogonal to each other. Due to the orthogonality of  the  vectors $\ve{m}$,  the  wavefunction  vanishes at the interface and  a  particle will not be able to cross it.  The  topological excitation of  the SO(3)  order parameter in two spatial dimensions  is a skyrmion.    In  its core   the SO(3)  order parameter    is given  by $\ve{m}_\mathrm{c}$   and  at  infinity by $-\ve{m}_\mathrm{c}$ . The  core is  surrounded by a vortex  in a  plane  perpendicular to  $\ve{m}_\mathrm{c}$  that   acts as  an infinite   potential  barrier.
As  a consequence,  we can foresee  that a skyrmion of the SO(3)   order  parameter  will  trap  charge in its  core.  Mean-calculations supporting this  point of  view are  presented in \cite{supplemental}.  Understanding if  this \textit{topological localization}  is essential  for the description of the  observed phase  transitions  remains an open issue.

The phase diagram of our model contains two ordered phases with low-lying Goldstone modes.  As mentioned above, it is possible to dope  our  system without encountering a negative-sign problem. Hence, numerical  simulations aimed at understanding the nature of the doping-induced   transitions to correlated metals (or superconductors~\cite{kozii19}) should be feasible; these will be subject of future work.


\begin{acknowledgments}
The authors gratefully acknowledge the Gauss Centre for Supercomputing e.V.\ (www.gauss-centre.eu) for funding this project by providing computing time on the GCS Supercomputer SUPERMUC-NG at Leibniz Supercomputing Centre (www.lrz.de).
This research has been supported by the Deutsche Forschungsgemeinschaft through the W\"urzburg-Dresden Cluster of Excellence on Complexity and Topology in Quantum Matter -- \textit{ct.qmat} (EXC 2147, Project No.\ 390858490), SFB 1170 on Topological and Correlated Electronics at Surfaces and Interfaces (Project No.\  258499086), SFB 1143 on Correlated Magnetism (Project No.\ 247310070), and the Emmy Noether Program (JA2306/4-1, Project No.\ 411750675)
\end{acknowledgments}

\bibliographystyle{shortapsrev4-2}
\bibliography{SO3}


\clearpage


\setcounter{figure}{0}
\setcounter{equation}{0}
\numberwithin{equation}{section}
\numberwithin{figure}{section}
\renewcommand\thefigure{S\arabic{figure}}
\renewcommand\theequation{S\arabic{equation}}

\title{Supplemental Material for ``Exotic quantum criticality in Dirac systems: Metallic and deconfined''}
\date{\today}
\maketitle

\section{Mean-field calculation}
In this section, we discuss the mean-field calculation of the lattice model~\eqref{eq:model}. In the main text, we have introduced two mean-field order parameters  for the   SO(3),  $m_{\alpha}/2 = (-1)^{\ve{i}} \langle c^{\dagger}_{\ve{i} \lambda} K^{\alpha} c^{\phantom\dagger}_{\ve{i} \lambda} \rangle $, and  the  U(1),   $V/2 = (-1)^{\ve{i}} \langle n^{\dagger}_{\ve{i}\sigma}  \rangle  $,  orders. On a bipartite lattice,  $(-1)^{\ve{i}}$  takes the  value $+1$ ($-1$)   on  sublattice  $A$ ($B$).
Starting from Eq.~\eqref{mean_field.eq}  of the main text,     our  mean-field  approximation reads: 
\begin{align}
-J\sum_{\ve{i}\alpha}\left(c_{\ve{i}}^{\dagger}K^{\alpha}\tau_{z}c_{\ve{i}}\right)^{2}\approx & -J\sum_{i\alpha\lambda}m_{\alpha}c_{\ve{i},\lambda}^{\dagger}K^{\alpha}c_{\ve{i},\lambda}\nonumber \\
 & -JV\sum_{\ve{i}\alpha\sigma\sigma^{\prime}}\left|\epsilon_{\alpha\sigma\sigma^{\prime}}\right|\left(n_{\ve{i}\sigma^{\prime}}+n_{\ve{i}\sigma}^{\dagger}\right)\nonumber \\
 & +J\sum_{\ve{i}\alpha}\left(V^{2}+\frac{\vec{m}^{2}}{2}\right).
 \label{eq:HStransform}
\end{align}
The  mean-field  Hamiltonian  consists  of the terms on the right-hand-side of the above equation, supplemented by the fermion hopping term.
To  proceed,   let us define  by $a_{\ve{R},\sigma,\lambda}^{\dagger}$ and $b_{\ve{R},\sigma,\lambda}^{\dagger}$  
 the  electron  creation operators  on  sublattices  $A$ and $B$, respectively,  in the unit cell $\ve{R}$.     After  Fourier  transformation to  
 momentum space, we  obtain:
\begin{align}
H_\text{MF}= & -t\sum_{\ve{k}\sigma\lambda}\left(f(k)a_{\ve{k}\sigma\lambda}^{\dagger}b^{}_{\ve{k}\sigma\lambda}+\text{h.c.}\right)+2J\sum_{\ve{k}\alpha}\left(V^{2}+\frac{m^{2}}{2}\right)\nonumber \\
 & -J\sum_{\ve{k}\alpha\lambda}m_{\alpha}\left(a_{\ve{k}\lambda}^{\dagger}K^{\alpha}a^{}_{\ve{k}\lambda}-b_{\ve{k}\lambda}^{\dagger}K^{\alpha}b^{}_{\ve{k}\lambda}\right)\nonumber \\
 & -JV\sum_{\ve{k}\alpha\sigma\sigma^{\prime}}\left|\epsilon_{\alpha\sigma\sigma^{\prime}}\right|\left[\left(n_{\ve{k}\sigma}^{a\dagger}-n_{\ve{k}\sigma}^{b\dagger}\right)+\left(n_{\ve{k}\sigma}^{a}-n_{\ve{k}\sigma}^{b}\right)\right]\nonumber \\
= & \sum_{\ve{k}}\hat{\psi}_{\ve{k}}^{\dagger}M_{\ve{k}}\hat{\psi}_{\ve{k}}+2J\sum_{\ve{k}\alpha}\left(V^{2}+\frac{\vec{m}^{2}}{2}\right),
\end{align}
where  we  have introduced   the   12-component   spinor  $\psi_{\ve{k}}^{\dagger}=\left( \{a_{\ve{k},\sigma,\lambda}^{\dagger}\},\{b_{\ve{k},\sigma,\lambda}^{\dagger}\}\right)$   and 
\begin{equation}
M_{\ve{k}}=\begin{bmatrix}M_{\ve{k}}^{aa} & -tf(\ve{k})\hat{I}\\
-tf^{*}(\ve{k})\hat{I} & M_{\ve{k}}^{bb}
\end{bmatrix}
\end{equation}
with 
 $f(\ve{k})=e^{-ik_{x}a}+e^{i(k_{x}/2)a+i(\sqrt{3}k_{y}/2)a}+e^{i(k_{x}/2)a-i(\sqrt{3}k_{y}/2)a}$ and
\begin{align}
M_{\ve{k}}^{aa} & = 
\begin{bmatrix} & iJm_{3} & -iJm_{2} & -2JV\\
-iJm_{3} &  & iJm_{1} &  & -2JV\\
iJm_{2} & -iJm_{1} &  &  &  & -2JV\\
-2JV &  &  &  & iJm_{3} & -iJm_{2}\\
 & -2JV &  & -iJm_{3} &  & iJm_{1}\\
 &  & -2JV & iJm_{2} & -iJm_{1}
\end{bmatrix}.
\end{align}
Finally, $M_{\ve{k}}^{bb} =-M_{\ve{k}}^{aa}$. 

The values of the mean-field order parameters are determined by self-consistent  conditions,  $\left\langle \partial H_{MF}/\partial V\right\rangle =\left\langle \partial H_{MF}/\partial m\right\rangle =0$,    that translate to:
\begin{gather}
m_{\alpha}=\frac{1}{2}\sum_{\lambda}\left(\left\langle a_{\ve{k},\lambda}^{\dagger}K^{\alpha}a_{\ve{k},\lambda}\right\rangle -\left\langle b_{\ve{k},\lambda}^{\dagger}K^{\alpha}b_{\ve{k},\lambda}\right\rangle \right), \\
V=\frac{1}{4}\sum_{\ve{k}\alpha\sigma\sigma^{\prime}}\left|\epsilon_{\alpha\sigma\sigma^{\prime}}\right|\left(\langle n_{\ve{k}\sigma}^{a\dagger}\rangle -\langle n_{\ve{k}\sigma}^{b\dagger}\rangle+\langle n_{\ve{k}\sigma}^{a}\rangle - \langle n_{\ve{k}\sigma}^{b}\rangle \right).
\end{gather}
The numerical solution    of the above  equations give rise  to the results shown in Fig.~\ref{fig:mftphasediagram} of the main text.

\section{Magnetic-flux insertion}
To reduce  finite-size effects,  we   follow  ideas put   forward in  Ref.~\cite{Assaad01} and  include in
our  simulations  a  single   flux   quantum  traversing the  lattice.    In particular, the    non-interacting 
part of the Hamiltonian   reads: 
\begin{equation}
H_{0}=-t\sum_{\left< \ve{i},\ve{j} \right> } \ve{c}^{\dagger}_{\ve{i}} e^{\frac{2\pi i}{\Phi_{0}}  \tau_z \int_{i}^{i+\delta}\boldsymbol{A}(\boldsymbol{l}) \cdot d\boldsymbol{l}} \ve{c}^{}_{\ve{j}}+h.c.  \, .
\end{equation}
Here,  $\Phi_{0}$ is the flux quantum, and $\boldsymbol{A}(\boldsymbol{l})=-B\left(y,0,0\right)$  in the Landau gauge, with 
$B =  \frac{\Phi_0} { \left|   \ve{L}_1  \times  \ve{L}_2  \right | }$.
 That is,  a  single  flux quantum  traverses the  
lattice, lying in the  $x$-$y$ plane  and being spanned  by the   vectors  $\ve{L}_1 =  L \ve{a}_1 $  and  $\ve{L}_2 =  L \ve{a}_2$.
Including $\tau_z$    preserves  the  time  reversal  symmetry, 
 defined as $T^{-1}   z  c^{\dagger}_{\ve{i} \sigma \lambda} T  =   \overline{z}  i \tau^{y}_{\lambda \lambda'}c^{\dagger}_{\ve{i} \sigma \lambda'}$,
required   to  avoid  
the  negative sign problem.   Further details of the  implementation of the flux can  be found in Ref.~\cite{ALF_v2}.

\begin{figure}[tbp]
\begin{centering}
\includegraphics[width=\columnwidth]{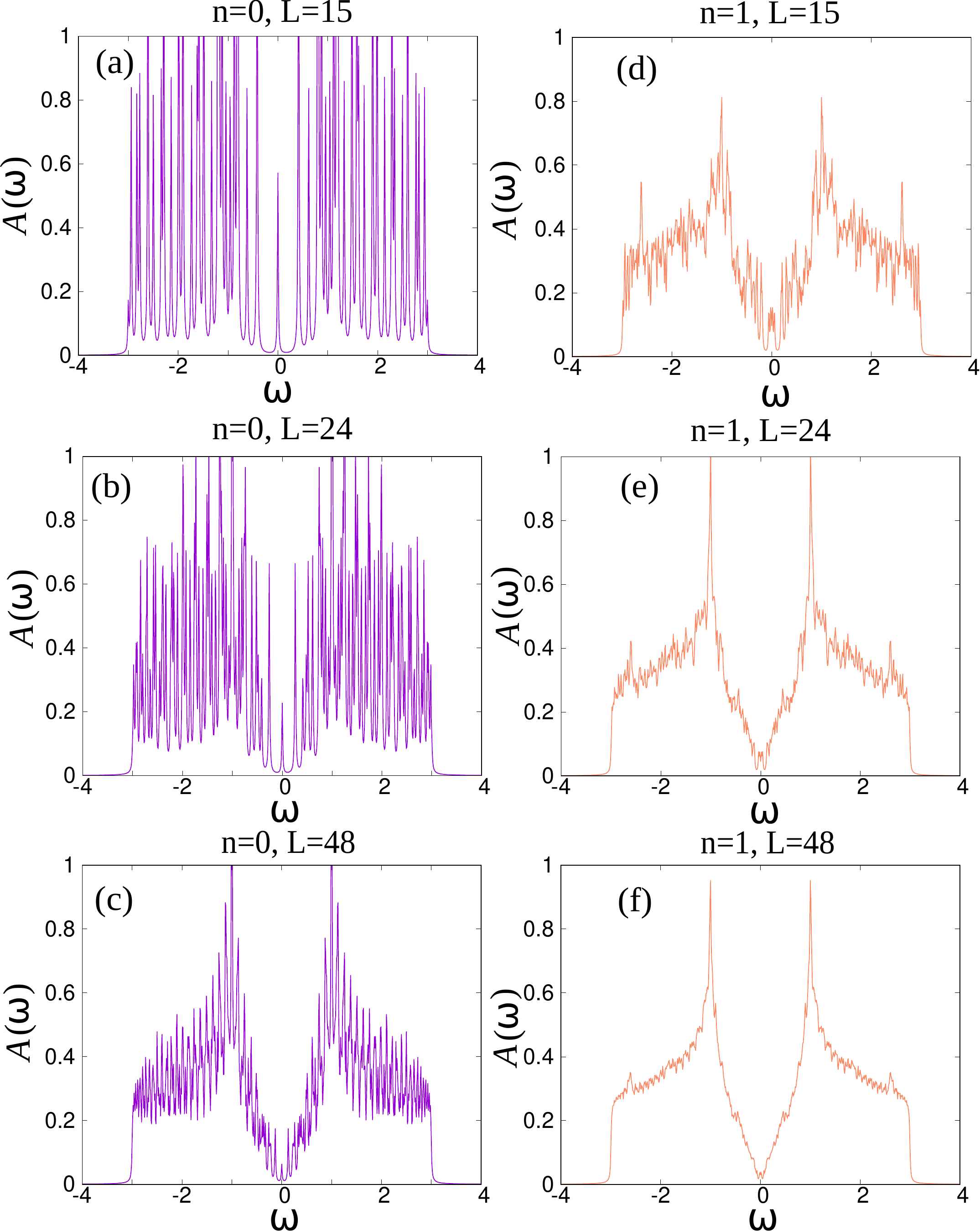}
\par\end{centering}
\caption{Local density of states $N(\omega)$ of  the tight-binding Hamiltonian $H_{0}$ 
(a)--(c) without magnetic flux insertion,  $B = 0$, and (d)--(f) with magnetic flux insertion, $B = \frac{\Phi_0} { \left|   \ve{L}_1  \times  \ve{L}_2  \right |}$, for different system sizes from $L=15$ (top) to $L=48$ (bottom).
For the plot, we have used a  broadening of  $\delta = 0.01t$  and  a  discretization of the 
frequency range of $ \Delta \omega = 0.01t$.}
\label{fig:LocalDos_test}
\end{figure}

By  construction,  the  magnetic field  vanishes in the thermodynamic limit.  
However   the convergence  of   observables
is   greatly  improved  when  including  the flux.   Here, we   exemplify  this by  using the   tight binding
Hamiltonian $H_{0}$ on the  honeycomb lattice. In Fig.~\ref{fig:LocalDos_test},
we present the finite-size fermion local density of states 
$N(\omega)=\frac{1}{\pi L^2 }\sum_{\ve{k}}\mathrm{Im}\,\text{Tr}\,G(\ve{k},\omega)$.
Panels (a)--(c)  and (d)--(f)  provide  a  comparison   $N(\omega)$   without and  with magnetic flux insertion,  respectively.  
As apparent, the  inclusion of the flux  improves the  quality of the density of states  on 
 finite-size lattices considerably.

\begin{figure}[tbp]
\begin{centering}
\includegraphics[width=\columnwidth]{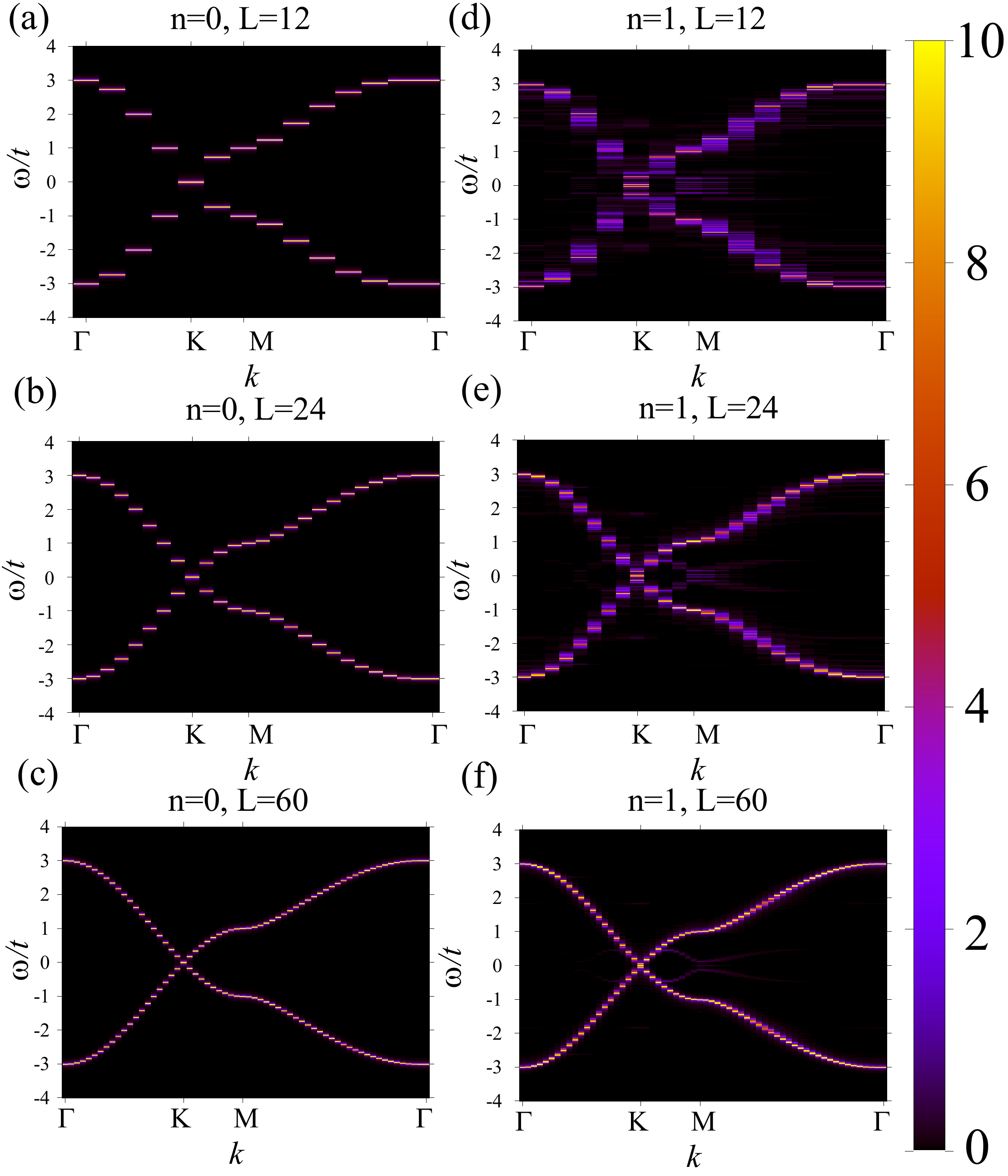}
\par\end{centering}
\caption{Fermion spectral function $A(k,\omega)$ for tight-binding Hamiltonian
$H_{0}$
(a)--(c) without magnetic flux insertion,  $B = 0$, and (d)--(f) with magnetic flux insertion, $B = \frac{\Phi_0} { \left|   \ve{L}_1  \times  \ve{L}_2  \right |}$, for different system sizes from $L=12$ (top) to $L=60$ (bottom).
\label{fig:akw_z_flux}}
\end{figure}

This   improvement, however, comes at a price,  namely  the  breaking  of 
translational  symmetry along $\ve{a}_1$   and  $\ve{a}_2$. 
In fact,  for our  choice of the  magnetic field,  the  magnetic unit  cell  
corresponds to the full lattice.   As a consequence,  the momentum $\ve{k}$ 
is no longer a good quantum number and  the fermion Green's function is not diagonal in
this quantity. 
Nevertheless, we can still
implement  the Fourier transformation  and  define the fermion spectral
function  as $G(\ve{k},\omega)=\frac{1}{L^2}\sum_{\boldsymbol{i},\boldsymbol{j}}e^{-i\boldsymbol{k}\cdot(\boldsymbol{i}-\boldsymbol{j})}
G(\boldsymbol{i},\boldsymbol{j},\omega)$.
In Panels (a)--(c)  and  (d)--(f) of Fig.~\ref{fig:akw_z_flux},   we plot the 
fermion spectral function $A(\ve{k},\omega)=\frac{1}{\pi}\mathrm{Im}\,\text{Tr}\,G(\ve{k},\omega)$  without  and 
with magnetic flux insertion, respectively.
One  observes  that in the presence of the flux,  $A(\ve{k},\omega)$  is  still dominated by the single-particle band
observed in the  absence of   flux.  
However,  novel features  stemming  from scattering  between  different momenta   emerge    at low energy and predominantly near the
the  high-symmetry point $\mathrm{M}=(0,{2\pi}/{\sqrt{3}})$. 
Clearly,  as a function of system size, these  features  become  weaker  and  ultimately 
disappear. 

\begin{figure}[tbp]
\begin{centering}
\includegraphics[width=.7\columnwidth]{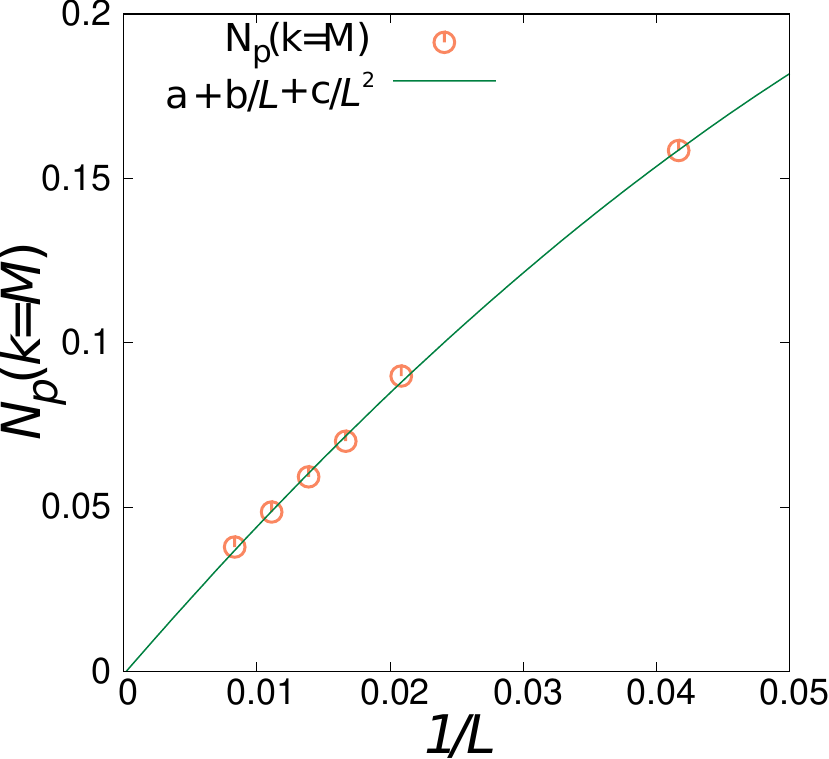}
\par\end{centering}
\caption{Fermion spectral weight $N_{p}$ at $\ve k = \mathrm{M}$, integrated over finite frequency range $\omega \in [-0.3,0.3]$, as function of $1/L$. The low-energy spectral weight is finite on finite-size lattices, but extrapolates to zero in the thermodynamic limit.
\label{fig:dos_vs_L}}
\end{figure}

To show  this explicitly, we study the quantity $N_{p}(\ve{k})=\int_{\omega_{1}}^{\omega_{2}}A(\ve{k},\omega)d\omega$,
which integrates  the spectral function at given momentum $\ve{k}$ over a finite frequency range.
We choose $\ve{k}=\mathrm{M}$ and set $\omega_{1}=-0.3$ and $\omega_{2}=0.3$,
corresponding to a  range   of  frequencies  located   well below the van Hove singularity at $\omega=\pm1$.
In Fig.~\ref{fig:dos_vs_L},
we   show  this quantity as a function of  inverse
system size $1/L$.
As  expected,  we  observe  that this  quantity  vanishes in the  thermodynamic  limit. 

\section{GN-SO(3) transition: Data-collapse analysis}

Dimensionless RG-invariant quantities  are  the tool  of choice to  investigate critical phenomena.  Here,  we  consider  the   quotient of correlation length $\xi$  to  the linear  system  size  $L$.    The correlation length  can be extracted from  the real-space  two-point correlation function of the order-parameter field,  $\widetilde{S}(\ve{r})$,   via 
\begin{align}
\xi^2=\frac{1}{2d}\frac{\sum_{\ve{r}}|\ve{r}|^2 \widetilde{S}(\ve{r})}{\sum_{\ve{r}}\widetilde{S}(\ve{r})},
\label{eq:xi_in_real}
\end{align}
where  $d$ corresponds to the spatial  dimension. 


Following Ref.~\onlinecite{toldin2015fermionic}, we  can use Eq.~(\ref{eq:xi_in_real}), to  define   the correlation length as
\begin{align}
\xi^2_{s,\kappa,\rho}(L)
&=\frac{\smashoperator[r]{\sum_{\substack{(-1+\kappa)L_1+1\le n_1\le \kappa L_1 \\(-1+\rho)L_2+1\le n_2\le \rho L_2}}}|n_1\ve{a}_1+n_2\ve{a}_2|^2S^\mathrm{SO(3)}(n_1\ve{a}_1+n_2\ve{a}_2)}{\smashoperator{\sum_{\substack{0\le n_1\le L_1-1 \\0\le n_2\le L_2-1}}}S^\mathrm{SO(3)}(n_1\ve{a}_1+n_2\ve{a}_2)}
\end{align}
on a  finite-size system. 
In the above,  $\ve{a}_1$ and $\ve{a}_2 $  correspond to the lattice  vectors of the  honeycomb lattice. We   consider  lattices   spanned  by the vectors  $L_1 \ve{a}_1 $  and $L_2 \ve{a}_2 $  with  periodic boundary  conditions. 
Since in our simulations $L \equiv  L_1 = L_2$  is a multiple of 3  we can choose  $\kappa=\rho=1/3$. In the following discussion, we denote $R^\mathrm{SO(3)}_{\xi}(L)=\xi_{s,\kappa=1/3,\rho=1/3}(L)/L$ as the RG-invariant quantity obtained in real space.

The correlation ratio \cite{Kaul15}  we discuss in the main text  is equally  an RG-invariant quantity  and  corresponds to an alternative  estimate of  $\xi^2/L^2$.    For  instance,   consider the Ornstein-Zernike form, $S^\mathrm{SO(3)}(\ve{k})\propto {A}/{(\xi^{-2}+|k|^2)}$
where $A$ is a constant.  With  this  ansatz, the  correlation ratio  reads: 
\begin{align}
R^\mathrm{SO(3)}_\mathrm{c}(L)  & \equiv  1-\frac{S^\mathrm{SO(3)}(\mathbf{Q}+d\ve{k})}{S^\mathrm{SO(3)}(\mathbf{Q})} \nonumber\\&= \frac{4\pi^2(\xi/L)^2}{1+4\pi^2(\xi/L)^2}=f(\xi/L)
\end{align}

\begin{figure}[tbp]
\centering
\includegraphics[width=\columnwidth]{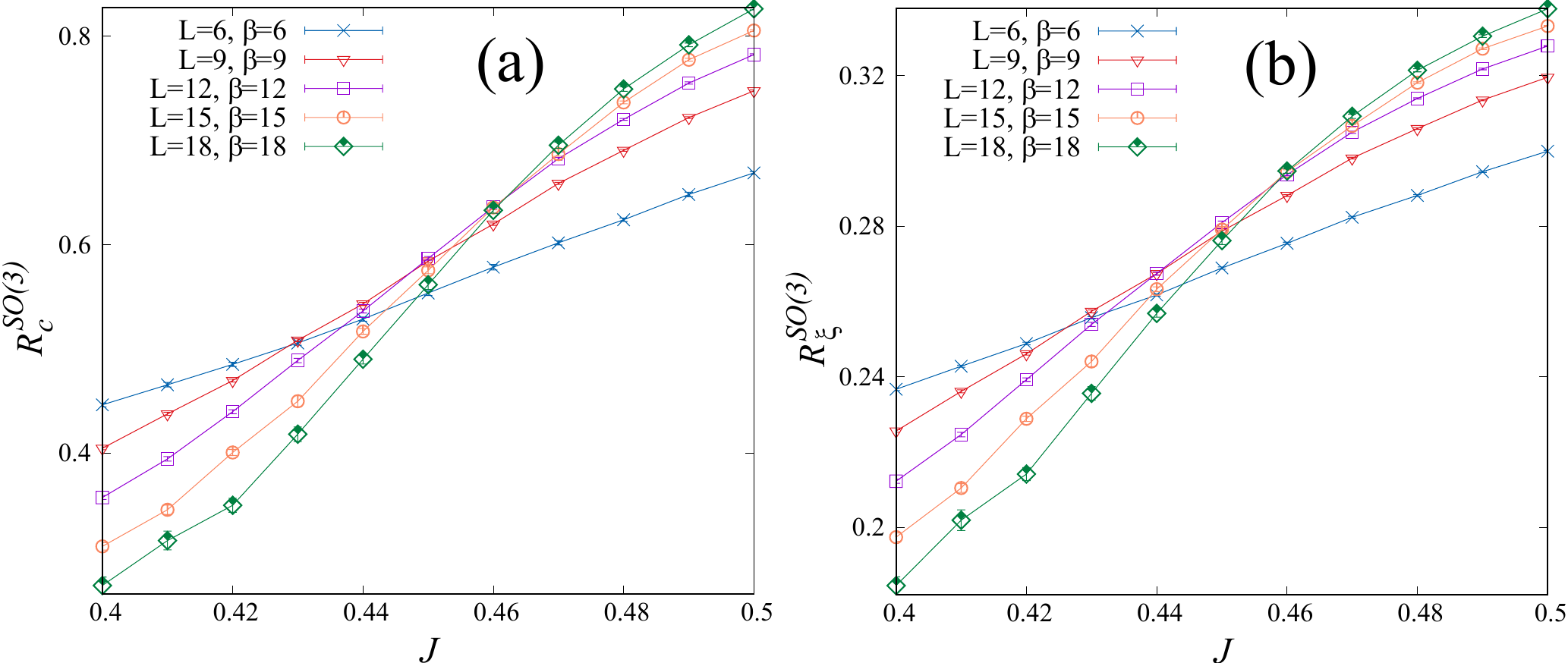}
\caption{RG-invariant quantities (a)~$R^\mathrm{SO(3)}_\mathrm{c}$ and (b)~$R^\mathrm{SO(3)}_{\xi}$ as function of coupling constant $J$ for different lattice sizes.}
\label{fig:rg_invariant}
\end{figure}

The two RG-invariant quantities obey the following scaling form  (for  an explicit proof  we refer the  reader  to Ref.~\cite{liu2021grossneveu})  close to the critical point 
\begin{align}
R_\mathrm{c}^\mathrm{SO(3)}(j,L) & =g_{1}(L^{z}/\beta,\left(J-J_\mathrm{c1}\right)L^{1/\nu},L^{-\omega})\nonumber\\
 & \approx f_{0}^{R}(jL^{1/\nu})+L^{-\omega}f_{1}^{R}(jL^{1/\nu})
\end{align}
\begin{align}
R_{\xi}^\mathrm{SO(3)}(j,L) & =g_{2}(L^{z}/\beta,\left(J-J_\mathrm{c1}\right)L^{1/\nu},L^{-\omega})\nonumber \\
 & \approx f_{2}^{R}(jL^{1/\nu})+L^{-\omega}f_{4}^{R}(jL^{1/\nu})
\end{align}
where $j =J-J_\mathrm{c1}$ and  $g_1$ and $g_2$ are scaling functions. Here, we assume that the  dynamical exponent $z=1$ and adopt a $\beta=L$ scaling. The term  $L^{-\omega}$, $\omega > 0$,  takes into account   leading corrections to scaling  that  vanish as $L\rightarrow\infty$.  For finite systems, corrections to scaling account  for  a drift in  the   crossing point of  RG-invariant quantities as a function of system  size.  In Fig.~\ref{fig:rg_invariant}, we show the RG-invariant quantities $R^\mathrm{SO(3)}_\mathrm{c}$ and  $R^\mathrm{SO(3)}_{\xi}$ for different lattice sizes $L$. Both  curves show a common crossing point at $J\approx 0.46$  for system sizes $L \ge 12$.  This observation suggests that  corrections  to scaling are weak in our FSS analysis and  that we can neglect them in our analysis of the  exponents. We perform a polynomial expansion of  $f_0^R(jL^{1/\nu})$  and $f_2^R(jL^{1/\nu})$ 
\begin{equation}
R_\mathrm{c}^\mathrm{SO(3)}(j,L)=\sum_{n=0}^{n_\text{max}}a_{n}j^{n}L^{n/\nu}
\end{equation}
\begin{equation}
R_{\xi}^\mathrm{SO(3)}(j,L)=\sum_{n=0}^{n_\text{max}}b_{n}j^{n}L^{n/\nu}
\end{equation}
and fit the data to these forms with  $\{a_n\}$, $\{b_n\}$, critical exponent $\nu$, and critical point $J_\mathrm{c}$ as  fit parameters.  Our  results are summarized in Tables~\ref{Rc_fit} and \ref{Rxi_fit}. We adjust the expansion order $n_\text{max}$ as  well as the  minimum size $L_\text{min}$  to take into account systematic errors and corrections to scaling.  
We also present the   $\chi^2/\text{DOF}$ value  of the fit, where $\text{DOF}$ refers to the number of degrees of freedom. In Table~\ref{Rc_fit},  $\chi^2/\text{DOF}$ takes  large values only for $L_\text{min}=6$ indicating the strong scaling correction for such small lattices. The best fit is obtained at $L_\text{min}=12$ and the  results have a weak dependence on the expansion order  for $n_\text{max}\ge 3$. 
Our results based on the  correlation ratio $R^\mathrm{SO(3)}_\mathrm{c}$  yield the estimates $J_\mathrm{c1}=0.461(1)$, $1/\nu=0.906(35)$. In Table~\ref{Rxi_fit}, the numerical  results based on the  RG-invariant quantity $R^\mathrm{SO(3)}_{\xi}$  are reported.  
The  results at fixed $L_\text{min}=12$ are independent on the expansion order $n_\text{max}$ and we find  consistent values,  $J_\mathrm{c1}=0.458(1)$, $1/\nu=0.937(21)$. 

\begin{table}[tbp]
    \caption{Results of fits of correlation ratio $R^\mathrm{SO(3)}_\mathrm{c}(L)$ close to $J_\mathrm{c1}$. $L_\text{min}$ is the minimum lattice size taken into account in the fits. $n_\text{max}$ is the polynomial expansion order  for  the scaling function.} 
    \centering 
    \def\arraystretch{1.5}
    \begin{tabular*}{\linewidth}{@{\extracolsep{\fill} } l l c c c}
    \hline\hline 
    \multicolumn{5}{c}{\hspace*{-3em}FSS analysis of $R^\mathrm{SO(3)}_\mathrm{c}(L)$ near $J_\mathrm{c1}$\hspace*{-3em}} \\
    \hline
    $L_\text{min}$ & $n_\text{max}$ & $J_\mathrm{c1}$ & $1/\nu$  & $\chi^2/\text{DOF}$ \\[0.5ex]
    \hline\hline
    \multirow{4}{*}{6} & 2 & 0.444(3) & 0.805(17) & 47.121 \\ 
	& 3 & 0.445(3) & 0.867(16) & 42.193 \\ 
	& 4 & 0.445(3) & 0.866(17) & 42.165 \\ 
	& 5 & 0.453(1) & 0.883(17) & 41.288\\ 
	\hline
	\multirow{4}{*}{9} & 2 & 0.453(1) & 0.827(18) & 8.936\\
	& 3 & 0.454(1) & 0.897(21) & 7.366 \\ 
	& 4 & 0.454(1) & 0.893(21) & 7.314\\ 
	& 5 & 0.454(1) & 0.894(21) & 7.276\\ 
	\hline
	\multirow{4}{*}{12} & 2 & 0.461(1) & 0.845(29) & 1.835 \\ 
	& 3 & 0.461(1) & 0.906(35) & 1.205\\ 
	& 4 & 0.461(1) & 0.906(35)  & 1.127\\ 
	& 5 & 0.461(1) & 0.901(37)  & 1.075\\ 
	\hline\hline
    \end{tabular*}
    \label{Rc_fit} 
\end{table}

\begin{table}[tbp]
    \caption{Results of fits of RG-invariant quantity $R^\mathrm{SO(3)}_{\xi}(L)$ close to $J_\mathrm{c1}$. $L_\text{min}$ is the minimum lattice size taken into account in the fits. $n_\text{max}$ is the polynomial expansion order for the scaling function.} 
    \centering 
    \def\arraystretch{1.5}
    \begin{tabular*}{\linewidth}{@{\extracolsep{\fill} } l l c c c}
    \hline\hline 
    \multicolumn{5}{c}{\hspace*{-3em}FSS analysis of $R^\mathrm{SO(3)}_{\xi}(L)$ near $J_\mathrm{c1}$\hspace*{-3em}} \\
    \hline
    $L_\text{min}$ & $n_\text{max}$ & $J_\mathrm{c1}$ & $1/\nu$  & $\chi^2/\text{DOF}$ \\[0.5ex]
    \hline\hline
    \multirow{4}{*}{6} & 2 & 0.436(4) & 0.894(9) & 99.826 \\ 
	& 3 & 0.435(4) & 0.852(11) & 98.463 \\ 
	& 4 & 0.437(4) & 0.918(12) & 91.740 \\ 
	& 5 & 0.437(4) & 0.905(12) & 91.486\\ 
	\hline
	\multirow{4}{*}{9} & 2 & 0.449(2) & 0.849(14) & 11.432\\
	& 3 & 0.449(2) & 0.850(13) & 11.373 \\ 
	& 4 & 0.449(2) & 0.872(15) & 10.662\\ 
	& 5 & 0.449(2) & 0.872(15) & 10.629\\ 
	\hline
	\multirow{4}{*}{12} & 2 & 0.458(1) & 0.937(21) & 2.909 \\ 
	& 3 & 0.457(1) & 0.924(23) & 2.762\\ 
	& 4 & 0.458(1) & 0.935(26)  & 1.754\\ 
	& 5 & 0.458(1) & 0.930(26)  & 1.684\\ 
	\hline\hline
    \end{tabular*}
    \label{Rxi_fit} 
\end{table}

For the analysis of the  bosonic, $\eta_{\phi}$, and  fermionic, $\eta_{\psi}$,  anomalous dimensions, we numerically fit the FSS behavior of the SO(3) order parameter  $m^2(j,L)$ and the quotient $Z(j,L)$.  In the proximity of the critical point, these two quantities  are expected to scale  as: 
\begin{align}
m^2(j,L) & =L^{2-d-z-\eta_{\phi}}g^m(L^{z}/\beta,\left(J-J_\mathrm{c1}\right)L^{1/\nu},L^{-\omega})\nonumber\\
 & \approx L^{-1-\eta_{\phi}}(f^{m}(jL^{1/\nu})+L^{-\omega}f_{1}^{m}(jL^{1/\nu}))\nonumber\\
 & =L^{-1-\eta_{\phi}}(\widetilde{f}^{m}(R_\mathrm{c/\xi}^\mathrm{SO(3)})+L^{-\omega}\widetilde{f}_{1}^{m}(R_\mathrm{c/\xi}^\mathrm{SO(3)}))
 \label{eq:FSS_m2}
\displaybreak[1]\\
Z(j,L) & =L^{-\eta_{\psi}}g^{z}(L^{z}/\beta,\left(J-J_\mathrm{c1}\right)L^{1/\nu},L^{-\omega})\nonumber \\
 & \approx L^{-\eta_{\psi}}(f^{z}(jL^{1/\nu})+L^{-\omega}f^{z}(jL^{1/\nu}))\nonumber\\
 & =L^{-\eta_{\psi}}(\widetilde{f}^{z}(R_\mathrm{c/\xi}^\mathrm{SO(3)})+L^{-\omega}\widetilde{f}^{z}(R_\mathrm{c/\xi}^\mathrm{SO(3)}))
  \label{eq:FSS_Zqp}.
\end{align}
In the  last  equalities of the  above equations and in an attempt  to minimize corrections to scaling,  we  have replaced $ jL^{1/\nu}$  by 
$R^\mathrm{SO(3)}_\mathrm{c}$ or $R^\mathrm{SO(3)}_{\xi}$.   For    $\beta = L $  scaling (appropriate for  $z=1$) and  in  the absence of corrections to scaling,  the correlation ratios are  functions of $ jL^{1/\nu} $.

\begin{figure}[tbp]
\centering
\includegraphics[width=\columnwidth]{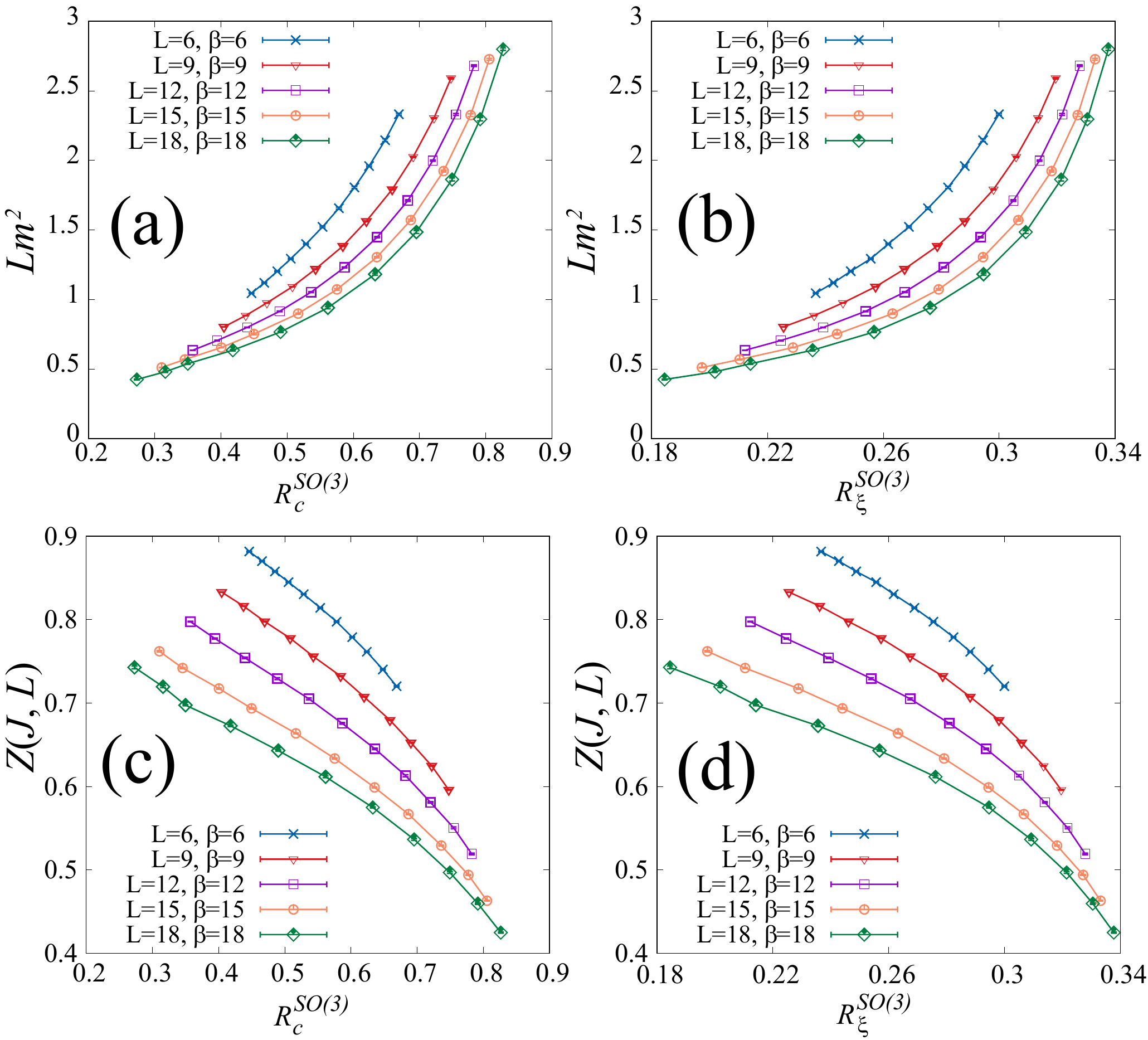}
\caption{FSS  behavior of  the  SO(3) order parameter $Lm^2$ and quantity $Z$ as a function of RG-invariant quantities $R^\mathrm{SO(3)}_\mathrm{c}$ and $R^\mathrm{SO(3)}_{\xi}$ in the proximity of the metallic transition point $J_\mathrm{c1}$. 
}
\label{fig:m2_Zqp_vs_R}
\end{figure}

In Fig.~\ref{fig:m2_Zqp_vs_R}, we show $Lm^2$ and $Z$ as a function of $R^\mathrm{SO(3)}_\mathrm{c}$ and $R^\mathrm{SO(3)}_{\xi}$.  The decrease of $Lm^2$ and $Z$ when increasing the system size $L$ are  a  consequence of the  anomalous dimensions $\eta_{\phi}$ and $\eta_{\psi}$ in Eqs.~(\ref{eq:FSS_m2}) and (\ref{eq:FSS_Zqp}).
Omitting corrections to scaling  and using a polynomial form for the scaling  functions,  we  can determine the anomalous dimensions $\eta_{\phi}$ and $\eta_{\psi}$.  Our results as function of $L_\text{min}$ and  of the maximal expansion order $n_\text{max}$ are listed in Tables~\ref{eta_fit_vs_Rc} and \ref{eta_fit_vs_Rxi}.  

In Table~\ref{eta_fit_vs_Rc}, we analyze  the data as a  function of $R_\mathrm{c}^\mathrm{SO(3)}$. The estimated value $\eta_{\phi}$ has small $\chi^2/\text{DOF}$, which indicates a good agreement with the scaling form. The fit to the data in Table~\ref{eta_fit_vs_Rc} yields a stable estimate of  $\eta_{\phi}=0.470(13)$   with respect to  $L_\text{min}$ and  to $n_\text{max}$.  The $\chi^2/\text{DOF}$ value of $\eta_{\psi}$ in Table~\ref{eta_fit_vs_Rc} shows a significant drop from $L_\text{min}=6$ to $L_\text{min}=9$, but  is stable when varying $L_\text{min}=9$ to $L_\text{min}=12$. The  dependence on the  expansion power $n_\text{max}$   has virtually no  effect on the estimated value of $\eta_{\psi}$.   Table~\ref{eta_fit_vs_Rc}   shows  that the  fermion anomalous dimension is  weakly   dependent on $L_\text{min}$ and  $n_\text{max}$ and  takes the value   $\eta_{\psi}=0.292(10)$. 

In Table~\ref{eta_fit_vs_Rxi},  we analyze the data in terms of the  RG invariant $R_{\xi}^\mathrm{SO(3)}$. The scaling analysis of $\eta_{\phi}$  is less favorable than that of  Table~\ref{eta_fit_vs_Rc}. The $\chi^2/\text{DOF}$  values  of $\eta_{\phi}$ in Table~\ref{eta_fit_vs_Rxi} are larger when $n_\text{max}<4$. Although the estimated values vary as a function $L_\text{min}$ and $n_\text{max}$, they are  consistent with estimated values in Table~\ref{eta_fit_vs_Rc}  when taking into account the statistical uncertainty. The  results of $\eta_{\psi}$ have a stable small  $\chi^2/\text{DOF}$ for $L_\text{min}>9$ and are not sensitive to the variation of $n_\text{max}$. The estimated value of $\eta_{\psi}$  in Table~\ref{eta_fit_vs_Rxi} is consistent with the estimated value in Table~\ref{eta_fit_vs_Rc} within statistical uncertainty.

\begin{table}[tbp]
    \caption{Results of fits of the SO(3) order parameter $m^2(j,L)$ and  $Z(j,L)$ as a function of  the  correlation ratio $R^\mathrm{SO(3)}_\mathrm{c}(L)$ close to $J_\mathrm{c1}$. $L_\text{min}$ is the minimum lattice size taken into account in the fits. $n_\text{max}$ is the polynomial expansion order of the scaling function.} 
    \centering 
    \def\arraystretch{1.5}
    \begin{tabular*}{\linewidth}{@{\extracolsep{\fill} } l l c c c c}
    \hline\hline 
    \multicolumn{6}{c}{\hspace*{-3em}FSS analysis of $m^2(j,L)$ and $Z(j,L)$ via $R^\mathrm{SO(3)}_\mathrm{c}(L)$ near $J_\mathrm{c1}$\hspace*{-3em}} \\
    \hline
    $L_\text{min}$ & $n_\text{max}$ & $\eta_{\phi}$  & $\chi^2/\text{DOF}$ & $\eta_{\psi}$ & $\chi^2/\text{DOF}$ \\[0.5ex]
    \hline\hline
    \multirow{4}{*}{6} & 2 & 0.464(11) & 24.471 & 0.247(5) & 119.198 \\ 
	& 3 & 0.465(8)  & 11.419 & 0.248(6) & 115.36 \\ 
	& 4 & 0.467(7) & 8.238 & 0.247(5) & 109.831 \\ 
	& 5 & 0.468(7) & 7.931 & 0.248(6) & 109.226\\ 
	\hline
	\multirow{4}{*}{9} & 2 & 0.449(16) & 25.179 & 0.284(4) & 19.34\\
	& 3 & 0.456(8) & 6.485 & 0.285(3) & 16.659\\ 
	& 4 & 0.461(7) & 4.644 & 0.285(4) & 16.482\\ 
	& 5 & 0.461(7) & 4.474 & 0.285(4) & 16.169\\ 
	\hline
	\multirow{4}{*}{12} & 2 & 0.457(35) & 27.542 & 0.292(10) & 21.329 \\ 
	& 3 & 0.459(16) & 5.33 & 0.292(9) & 15.847\\ 
	& 4 & 0.470(13) & 3.21  & 0.292(9) & 15.614\\ 
	& 5 & 0.470(13) & 3.039  & 0.292(10) & 15.045\\ 
	\hline\hline
    \end{tabular*}
    \label{eta_fit_vs_Rc} 
\end{table}

\begin{table}[tbp]
    \caption{Results of fits of SO(3)  order parameter $m^2(j,L)$ and  $Z(j,L)$ as a function of RG-invariant quantity $R^\mathrm{SO(3)}_{\xi}(L)$ close to $J_\mathrm{c1}$. $L_\text{min}$ is the minimum lattice size taken into account in the fits. $n_\text{max}$ is the polynomial expansion order of the scaling function.} 
    \centering 
    \def\arraystretch{1.5}
    \begin{tabular*}{\linewidth}{@{\extracolsep{\fill}} l l c c c c}
    \hline\hline 
    \multicolumn{6}{c}{{\hspace*{-3em}FSS analysis of $m^2(j,L)$ and $Z(j,L)$ via $R^\mathrm{SO(3)}_{\xi}(L)$ near $J_\mathrm{c1}$\hspace*{-3em}}} \\
    \hline
    $L_\text{min}$ & $n_\text{max}$ & $\eta_{\phi}$  & $\chi^2/\text{DOF}$ & $\eta_{\psi}$ & $\chi^2/\text{DOF}$ \\[0.5ex]
    \hline\hline
    \multirow{4}{*}{6} & 2 & 0.451(31) & 374.096 & 0.233(5) & 115.704 \\ 
	& 3 & 0.482(15)  & 85.585 & 0.235(5) & 102.444 \\ 
	& 4 & 0.49(1) & 36.467 & 0.234(5) & 93.719 \\ 
	& 5 & 0.500(8) & 19.098 & 0.234(5) & 93.321\\ 
	\hline
	\multirow{4}{*}{9} & 2 & 0.435(49) & 459.103 & 0.268(4) & 27.286\\
	& 3 & 0.458(22) & 88.106 & 0.269(3) & 15.824\\ 
	& 4 & 0.486(12) & 22.402 & 0.268(3) & 15.443\\ 
	& 5 & 0.503(7) & 8.291 & 0.268(3) & 15.147\\ 
	\hline
	\multirow{4}{*}{12} & 2 & 0.44(1) & 478.842 & 0.278(12) & 29.888 \\ 
	& 3 & 0.433(44) & 81.388 & 0.278(9) & 14.69\\ 
	& 4 & 0.488(21) & 16.614 & 0.277(9) & 14.412\\ 
	& 5 & 0.506(13) & 5.886  & 0.277(9) & 15.045\\ 
	\hline\hline
    \end{tabular*}
    \label{eta_fit_vs_Rxi} 
\end{table}

\section{GN-SO(3) transition: Crossing-point analysis}

In this section, we provide a consistency check of the above-estimated exponents using crossing points.  The crossing-point analysis is a general and reliable way to extract the critical behavior \cite{shao2016quantum}. The crossing point with size increment $c$    is  defined as the coupling $j_{c,R}(L)$  that satisfies  $R^\mathrm{SO(3)}_\mathrm{c}(j_{c,R}(L),L)=R^\mathrm{SO(3)}_\mathrm{c}(j_{c,R}(L),L+c)$.  As  $L\rightarrow \infty$, the finite-size critical coupling $j_{c,R}(L)$ approches the critical coupling $J_\mathrm{c}$ as 
\begin{equation}
j_{c,R}(L)=J_\mathrm{c}+AL^{-e}
\end{equation}
where $e=1/{\nu}+\omega$ and $A$ is a nonuniversal constant \cite{toldin2015fermionic}. Numerically,  $j_{c,R}(L)$ can be estimated  using a   polynomial interpolation of the RG-invariant curves. Once  we have estimated the  finite-size critical coupling $j_{c,R}(L)$, it is straightforward to calculate the finite-size critical exponents using
\begin{gather}
\frac{1}{\nu(L,L+c)}=\frac{1}{\ln(r)} \ln \left\{ \frac{s(j_{c,R}(L),L+c)}{s(j_{c,R}(L),L)} \right\},
\\
\eta_{\phi}(L,L+c)=-\frac{1}{\ln(r)}\ln \left\{ \frac{m^2(j_{c,R}(L),L+c)}{m^2(j_{c,R}(L),L)} \right\} -1,
\\
\eta_{\psi}(L,L+c)=-\frac{1}{\ln(r)}\ln \left\{ \frac{Z(j_{c,R}(L),L+c)}{Z(j_{c,R}(L),L)} \right\},
\end{gather}
where $s(j,L)=\frac{d R^\mathrm{SO(3)}_\mathrm{c}(j,L)}{dJ}$ is the first-order derivative of the RG-invariant curve and $r=(L+c)/L$. The finite-size critical exponents discussed above scale to the  correct exponent with a rate controlled by the leading-correction-to-scaling exponent~$\omega$,
\begin{gather}
\frac{1}{\nu(L,L+c)}=\frac{1}{\nu}+dL^{-\omega},
\\
\eta_{\phi}(L,L+c)=\eta_{\phi}+gL^{-\omega},
\\
\eta_{\psi}(L,L+c)=\eta_{\psi}+kL^{-\omega}.
\end{gather}

\begin{figure}[tbp]
\centering
\includegraphics[width=\columnwidth]{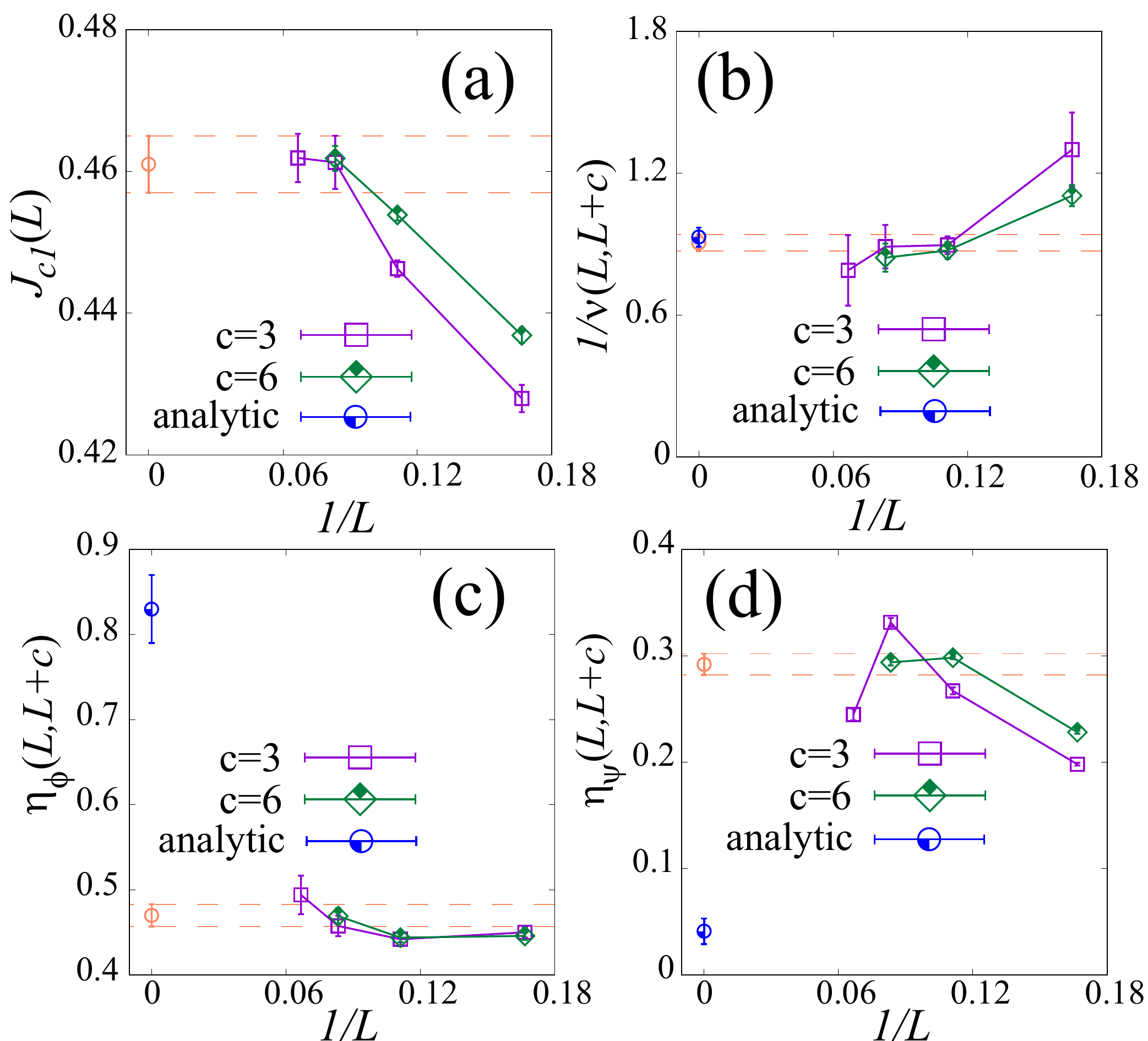}
\caption{System-size dependence of (a) critical coupling $J_\mathrm{c1}$, (b) critical exponent $1/\nu$, (c) SO(3)-order-parameter anomalous dimension $\eta_{\phi}$, and (d) fermion anomalous dimension $\eta_{\psi}$ at the GN-SO(3) transition, as obtained from the crossing-point analysis.
For comparison, the orange symbols and dotted lines in each panel indicate the estimates and statistical uncertainties, respectively, from the data-collapse analysis.
The blue dots in (b-d) indicate the corresponding estimates from the analytical calculations of~\cite{ray2021fractionalized}.}
\label{fig:Cross_Jc1}
\end{figure}

In Fig.~\ref{fig:Cross_Jc1}, we summarize the crossing-point analysis of the critical coupling and the critical exponents, and compare with the result obtained from the data-collapse analysis. We consider two different size increments $c=3$ and $c=6$, with $r=(L+3)/L$ and $r=(L+6)/6$, respectively. Our data quality  and size limitations  hinder extrapolation to the thermodynamic limit.  However, as shown in Fig.~\ref{fig:Cross_Jc1}(a),  the last two crossing points for $c=3$ are very close to each other and to the estimate from the data-collapse analysis, suggesting  weak   corrections to scaling in the correlation ratio $R^\mathrm{SO(3)}_\mathrm{c}$ for $L\ge 12$. In Fig.~\ref{fig:Cross_Jc1}(b), the finite size exponent $1/\nu(L,L+c)$ has  large error  bars that stem  from the difficulty to compute the first-order derivative of the RG-invariant curve. As shown in Fig.~\ref{fig:Cross_Jc1}(c), $\eta_{\phi}(L,L+c)$ shows weak increase as a  function of system size. Within our estimated error bars, the largest size $\eta_{\phi}(L,L+c)$ is consistent  with the  value obtained from data collapse. In Fig.~\ref{fig:Cross_Jc1}(d), $\eta_{\psi}(L,L+c)$ shows a smooth variation at $c=6$ and the finite-size exponent again matches the estimate from the data-collapse analysis.

\section{SO(3)-U(1) transition: Data-collapse analysis}

Our QMC data suggest the possibility of a second quantum critical point between the SO(3) semimetal and the U(1) insulator. If this is the case, it is worthwhile to estimate the critical exponents for this transition as well. In particular, the correlation-length exponent $\nu$, as well as the critical coupling $J_\mathrm{c2}$, are both expected to yield unique values when estimated from the two opposite sides of the transition, e.g., by using either the SO(3) or the U(1) order parameter. Following the same procedure as for the GN-SO(3) transition at $J_\mathrm{c1}$, we extract the critical exponents and  critical couplings  from the two different correlation ratios
\begin{align}
R_\mathrm{c}^\mathrm{SO(3)}(j,L) & =g_{1}(L^{z}/\beta,\left(J-J_\mathrm{c2}\right)L^{1/\nu},L^{-\omega})\nonumber \\
 & \approx f_{0}^{R}(jL^{1/\nu})=\sum_{n=0}^{n_\text{max}}a_{n}j^{n}L^{n/\nu}
\end{align}
and
\begin{align}
R_\mathrm{c}^\mathrm{U(1)}(j,L) & =g_{1}(L^{z}/\beta,\left(J-J_\mathrm{c2}\right)L^{1/\nu},L^{-\omega})\nonumber \\
 & \approx\tilde{f}_{0}^{R}(jL^{1/\nu})+L^{-\omega}\tilde{f}_{1}^{R}(jL^{1/\nu})\nonumber \\
 & =\sum_{n=0}^{n_\text{max}}a_{n}j^{n}L^{n/\nu}+L^{-\omega}\sum_{m=0}^{m_\text{max}}b_{m}j^{m}L^{m/\nu}.
\end{align}

The U(1) correlation curve shows a systematic drift of the crossing point  that  reflects the presence of a large leading-correction-to-scaling term $L^{-\omega}\tilde{f}_{1}^{R}$, see Fig.~\ref{fig:u1_so3_t0.pdf}(b) of the main text. In our analysis we  hence  took this term into account. Our results are summarized in Tables~\ref{Rc_fit_t0_so3} and \ref{Rc_fit_t0_u1}. In Table~\ref{Rc_fit_t0_so3}, the $\chi^2/\text{DOF}$ of the fit converge for $n_\text{max}>2$. The estimation of the critical point $J_\mathrm{c2}$ is stable and consistent with the crossing-point analysis.  The estimated exponent $1/\nu$ is not sensitive to  system size and we  estimate $1/\nu=1.673(58)$.

In Table~\ref{Rc_fit_t0_u1}, the error bars of the estimated exponent are larger due  to the  presence of the  leading-correction-to-scaling  term. Here we fix the expansion order $n_\text{max}=4$ and tune $m_\text{max}$   so as  to improve the fitting quality. 
The  quality of the fit is  mildly  improved   with growing values of $m_\text{max}$. 
 Again, the estimated $J_\mathrm{c2}$ is stable and consistent with the crossing-point analysis. The estimated exponent $1/\nu$ converge to a stable value for $m_\text{max}>1$. In conclusion, the numerical fitting of $R_\mathrm{c}^\mathrm{U(1)}(j,L)$ yield an estimation $1/\nu=1.458(642)$, that is  consistent with the estimate obtained from  $R_\mathrm{c}^\mathrm{SO(3)}(j,L)$, but with a larger error bar.     Importantly, the  two  estimated values of $J_\mathrm{c2} $ match well.
 
\begin{table}[tbp]
    \caption{Results of fits of correlation ratio $R^\mathrm{SO(3)}_\mathrm{c}(L)$ close to $J_\mathrm{c2}$. $L_\text{min}$ is the minimum lattice size taken into account in the fits. $n_\text{max}$ is the polynomial expansion order of  the scaling function.} 
    \centering 
    \def\arraystretch{1.5}
    \begin{tabular*}{\linewidth}{@{\extracolsep{\fill} } l l c c c}
    \hline\hline 
    \multicolumn{5}{c}{FSS analysis of $R^\mathrm{SO(3)}_\mathrm{c}(L)$ near $J_\mathrm{c2}$} \\
    \hline
    $L_\text{min}$ & $n_\text{max}$ & $J_\mathrm{c2}$ & $1/\nu$  & $\chi^2/\text{DOF}$ \\[0.5ex]
    \hline\hline
    \multirow{3}{*}{6} & 2 & 1.001(2) & 1.274(13) & 62.649 \\ 
	& 3 & 0.998(1) & 1.732(18) & 22.989 \\ 
	& 4 & 0.997(1) & 1.710(18) & 19.812 \\ 
	\hline
	\multirow{3}{*}{9} & 2 & 1.007(1) & 0.827(18) & 66.139\\
	& 3 & 1.001(2) & 1.706(29) & 19.803 \\ 
	& 4 & 1.000(2) & 1.650(32) & 16.797\\ 
	\hline
	\multirow{3}{*}{12} & 2 & 1.011(15) & 0.701(37) & 74.78 \\ 
	& 3 & 1.002(3) & 1.692(60) & 23.046\\ 
	& 4 & 1.001(3) & 1.673(58) & 20.249\\ 
	\hline\hline
    \end{tabular*}
    \label{Rc_fit_t0_so3} 
\end{table}

\begin{table}[tbp]
    \caption{Results of fits of correlation ratio $R^\mathrm{U(1)}_\mathrm{c}(L)$ close to $J_\mathrm{c2}$. $L_\text{min}$ is the minimum lattice size taken into account in the fits. $m_\text{max}$ is the polynomial expansion order of the leading correction scaling function. The main scaling function expansion order is fixed to $n=4$.} 
    \centering 
    \def\arraystretch{1.5}
    \begin{tabular*}{\linewidth}{@{\extracolsep{\fill} } l l c c c c}
    \hline\hline 
    \multicolumn{6}{c}{FSS analysis of $R^\mathrm{U(1)}_\mathrm{c}(L)$ near $J_\mathrm{c2}$} \\
    \hline
    $L_\text{min}$ & $m_\text{max}$ & $J_\mathrm{c2}$ & $1/\nu$  & $\omega$ & $\chi^2/\text{DOF}$ \\[0.5ex]
    \hline\hline
    \multirow{3}{*}{6} & 0 & 1.024(7) & 0.786(120) & 0.610(79) & 26.311 \\ 
	& 1 & 1.014(3) & 1.241(32) & 0.699(22) & 19.219 \\ 
	& 2 & 1.014(4) & 1.502(321) & 0.811(119) & 16.88 \\ 
	\hline
	\multirow{3}{*}{9} & 0 & 1.022(6) & 0.841(19) & 0.525(55) & 28.559\\
	& 1 & 1.014(3) & 1.309(36) & 0.582(60) & 22.7 \\ 
	& 2 & 1.011(5) & 1.371(403) & 0.824(135) & 20.2\\ 
	\hline
	\multirow{3}{*}{12} & 0 & 1.022(9) & 0.800(26) & 0.441(167) & 27.162 \\ 
	& 1 & 1.017(12) & 1.096(202) & 0.29(20) & 25.88\\ 
	& 2 & 1.007(17) & 1.458(642) & 0.756(165) & 15.662\\ 
	\hline\hline
    \end{tabular*}
    \label{Rc_fit_t0_u1} 
\end{table}

\section{SO(3)-U(1) transition: Correlation length}

\begin{figure}[tb]
\centering
\includegraphics[width=\columnwidth]{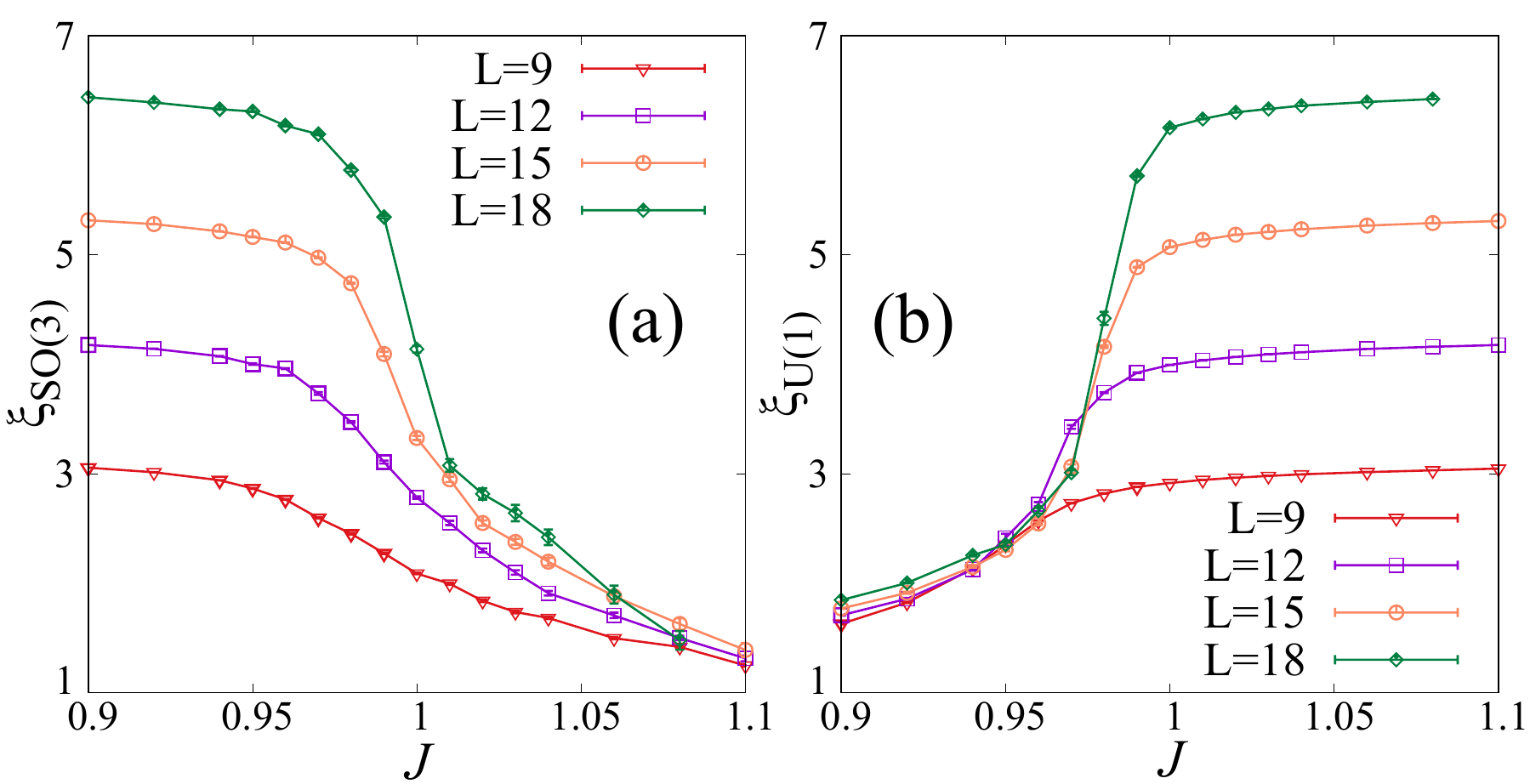}
\caption{Real-space correlation length $\xi$ as function of $J$ near the SO(3)-U(1) transition at  $J_\mathrm{c2} =1.007(17)$, computed from Eq.~(\ref{eq:xi_in_real})  using (a)  the SO(3)  and (b) the U(1) two-point correlators.}
\label{fig:Jc2_xi}
\end{figure}

In Fig.~\ref{fig:Jc2_xi}, we present the real-space correlation length $\xi_\mathrm{SO(3)/U(1)}$ near the SO(3)-U(1) transition at $J_\mathrm{c2}$ as obtained from the SO(3) and U(1) order parameters. We  extract   this  quantity from the two-point correlation function   following  Eq.~(\ref{eq:xi_in_real}). To be more precise,  consider the  U(1)  correlation  length in the  SO(3)-ordered  state corresponding to $J <  J_\mathrm{c2} = 1.007(17) $, Fig.~\ref{fig:Jc2_xi}(b).
At  a  first-order  transition, this quantity should saturate for increasing system size. At a   continuous  transition, it  should  be unbounded as $J$  approaches the critical point from below, $J \nearrow J_\mathrm{c2}$.  For  the accessible lattice sizes the  data of  Fig.~\ref{fig:Jc2_xi}(b)   is consistent  with the interpretation of a continuous transition.  

Similarly, Fig.~\ref{fig:Jc2_xi}(a)   shows   the SO(3)  correlation length as function of $J$. For a  continuous transition, we  expect this quantity to be unbounded in the U(1)   ordered phase upon approaching the critical point from above, $J \searrow J_\mathrm{c2}$.   In this case,  the data is  harder  to interpret,  since  the simulations on our largest lattice size, $L=18$, could hint towards a  saturation. 

\section{Topological defects of SO(3) order parameter}

In this section,  we  address the  question if  topological defects of the SO(3)  order parameter,  skyrmions,  can 
 localize  electrons   and  thereby  be important in  understanding  the criticality  between the  SO(3) semimetal and the U(1) insulator.  In order to do so,  we  consider  for an arbitrary   static configuration of the  SO(3)   order parameter, $N_{\alpha}\left(\mathbf{r}_{i}\right) $,  the mean-field Hamiltonian  
 \begin{align}
H = & -t\sum_{\left\langle \ve{i} \ve{j} \right\rangle \sigma}\left(c_{\ve{i}\sigma}^{\dagger}c^{}_{\ve{j}\sigma}+\text{h.c.}\right)
\nonumber \\&\quad 
 +\sum_{\ve{i}\sigma}(-1)^{\tau_{\ve{i}}}N_{\alpha}\left(\mathbf{r}_{i}\right)c_{\ve{i}\sigma}^{\dagger}K_{\sigma\sigma^{\prime}}^{\alpha}c^{}_{\ve{i}\sigma^{\prime}}
 \label{eq:skyrmion_mft}
\end{align}
on the  honeycomb lattice.  Here, $\ve{i}$  denotes a lattice  site  belonging to the unit-cell vector $\mathbf{r}_i$, which is taken to be the midpoint between the orbitals in the unit cell, $\tau_{\ve{i}}=1,2$ correspond to the   sublattice  index,
and $\sigma=1,2,3$ corresponds to the  SO(3)-spin-component index. 
To test  our  conjecture,  we ignore the layer degree of  freedom. 

\begin{figure}[tbp]
\centering
\includegraphics[width=\columnwidth]{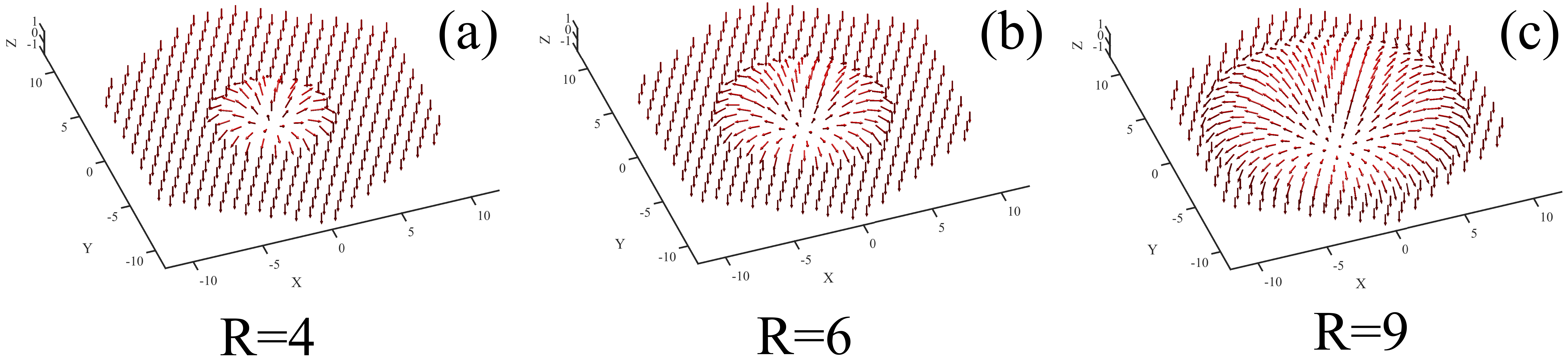}
\caption{Single-skyrmion configuration on $L=21$ honeycomb lattice with skyrmion radius (a) $R=4$, (b) $R=6$, and (c) $R=9$.}
\label{fig:Skyrmion}
\end{figure}

In the presence of periodic boundary conditions, we can consider  the following skyrmion configuration 
\begin{align}
\ve{N}(\mathbf{r}_{i})= & m_{0}\left[\sin\theta(\mathbf{r}_{i})\cos \phi(\mathbf{r}_{i}),\sin \theta(\mathbf{r}_{i}) \sin \phi(\mathbf{r}_{i}),
\cos \theta(\mathbf{r}_{i})\right]
\end{align}
with 
\begin{equation}
\theta(\mathbf{r}_{i})=\begin{cases}
2\arcsin\left(r_{i}/R\right) & \text{for } r_{i}<R,\\
\pi & \text{for } r_{i}>R,
\end{cases}
\end{equation}
where $\phi(\mathbf{r}_i)$ is the azimuthal angle at position $\mathbf{r}_i$, and $m_0$ is the amplitude of the SO(3) order parameter. The radius of the skyrmion  is determined by the parameter  $R$. In  Fig.~\ref{fig:Skyrmion}, we present real-space skyrmion configurations for different values of $R$ on an $L=21$ lattice. 

\begin{figure}[tbp]
\centering
\includegraphics[width=\columnwidth]{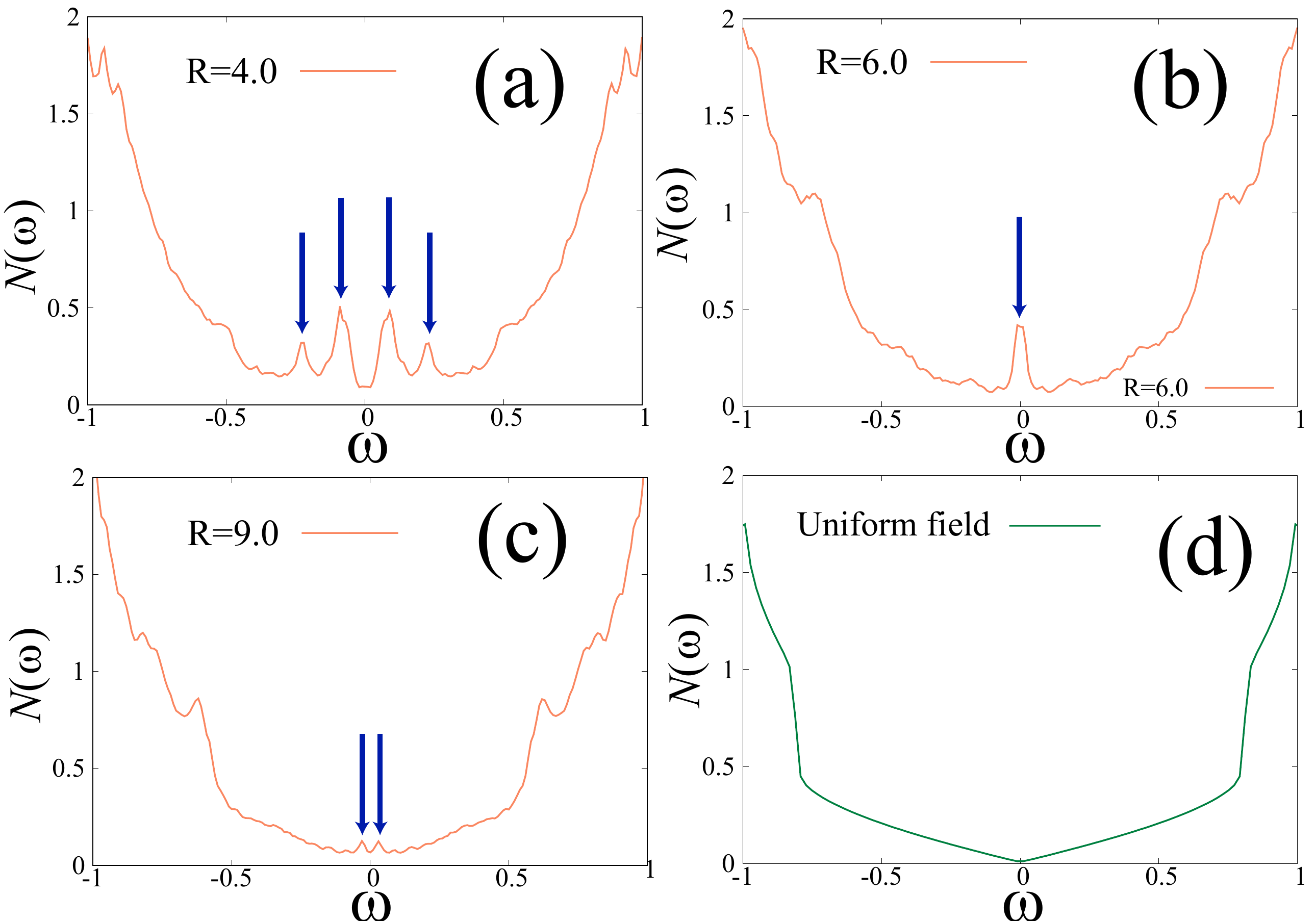}
\caption{Fermion local density of states at the center of the skyrmion with radius (a) $R=4$, (b) $R=6$, and (c) $R=9$, using $L=90$.  The blue arrows point out the discrete-energy-level feature close to the Fermi energy.
In (d), we plot the local density of states for a uniform field configuration $\ve{N}(\mathbf{r}_{i})=m_{0}\left(0,0,1\right)$ for comparison.}
\label{fig:fdos_Skyrmion}
\end{figure}

Our aim is to investigate   the impact of the skyrmion   on the  local  density of states (LDOS).  In order to do so, we diagonalize the Hamiltonian of  Eq.~(\ref{eq:skyrmion_mft}) on the honeycomb lattice and choose  $L=90$ so as to reduce  finite-size effects.   We  furthermore set $t=1$ and $m_0=1$.  In  Fig.~\ref{fig:fdos_Skyrmion},   we  plot  the LDOS, $N(\omega)$, at the center of  the skyrmion.      For small values of $R$,   we observe  distinct peaks in the  LDOS.   As   $R$ grows, they become less dominant  and  ultimately,   the  LDOS   converges to   that of the uniform field, Fig.~\ref{fig:fdos_Skyrmion}(d).   The  data is hence consistent    with the emergence of a  discrete  spectrum,  reminiscent of a  particle in a box,  generated by the   skyrmion.     The  step  from this \textit{topological localization}   to  a  theory of the phase  transition   necessarily involves the layer index.    Here, we can conjecture that skymions on different layers    bind,  and  that the localized states on the  respective  layers  hybridize to form the U(1)  order parameter.

\end{document}